\documentclass[a4paper,11pt]{article}
\usepackage{jheppub} 
\usepackage[T1]{fontenc} 

 \newcommand{\beq}[1]{\begin{equation}\label{#1}}
 \newcommand{\eeq}{\end{equation}}
 \newcommand{\bea}[1]{\begin{eqnarray}\label{#1}}
 \newcommand{\eea}{\end{eqnarray}}

\title{\boldmath Complexity, chaos and the moving $D_3$-brane}
 \author{Ai-chen Li }
\affiliation[a]{Institut de Ci\`encies del Cosmos, Universitat de Barcelona, Mart\'i i Franqu\`es 1, 08028 Barcelona, Spain}
\affiliation[b]{Departament de F\'isica Qu\`antica i Astrof\'isica, Facultat de F\'isica, Universitat de Barcelona, Mart\'i i Franqu\`es 1, 08028 Barcelona, Spain}
\emailAdd{aichenli@icc.ub.edu}

\abstract{We use the wave-function method developed in area of quantum information to investigate the quantum circuit complexity of the small quantum fluctuations around the probe $D_3$ brane moving in $AdS_5\times S^5$ bulk. In our consideration, the reference and target states are chosen as the vacuum state and the squeezed quantum state respectively. The evolution of parameters characterizing the squeezed quantum state are governed by the time-dependent $Schr\ddot{o}dinger$ equation, in which the Hamiltonian operator is derived from the perturbative action of $D_3$ brane. For a quantum chaotic system, some recent works indicate that the evolution of quantum circuit complexity could provide equivalent information like the out-of-time-order correlators. Basing on this inference, our results show that the quantum fluctuations around the non-BPS brane manifestly evolve into the chaotic regime at the late time, while the chaotic behavior is not easy to observe in case of BPS brane. In holographic viewpoint, it implies that the thermodynamic system consist of the $N=4$ supersymmetric particles in non-BPS states evolve into a chaotic system more easily than the one in BPS state.}


\begin{document} 
\maketitle
\flushbottom

\section{Introduction}

In recent years, theoretical physicists take a major step toward associating quantum information theory with gravity via the Anti-de Sitter/Conformal field theory (AdS/CFT) correspondence \cite{Maldacena:1997re,Witten:1998qj,Gubser:1998bc}. As a milestone in this research areas, the proposition of holographic entanglement entropy (HEE) \cite{Ryu:2006bv} provide a way to understand the entanglement entropy of the boundary CFT from the viewpoint of spacetime geometry in bulk. However, \cite{Hartman:2013qma} finds that the late-time growth behavior of the size of the Einstein-Rosen bridge (ERB) behind the horizon of an eternal AdS black hole can not be captured by the entanglement entropy of a thermo-field double (TFD) state after reaching the thermal equilibrium. To describe this late-time growth behavior of size of ERB, another physical quantity called quantum circuit complexity has been involved into the AdS/CFT dictionary. In quantum information theory, the quantum circuit complexity is identified with the minimum number of unitray operators that are used to evolve into the expected target state from an appropriate reference state. Specifically, there exists two main methods from viewpoint of holographic proposals in describing the evolution of quantum circuit complexity between reference state and target state on AdS boundary. One is known as complexity-volume (CV) conjecture \cite{Stanford:2014jda, Susskind:2014rva}, which supposes that the quantum circuit complexity on boundary is dual to the maximum volume of the ERB in bulk spacetime. Another version is complexity-action (CA) conjecture \cite{Brown:2015lvg, Brown:2015bva}, which associates the complexity on boundary to the gravitational action evaluated on a region of Wheeler-DeWitt patch in the bulk spacetime. Since then, many interesting and profound studies have been explored in term of the CV and CA conjectures of holographic complexity \cite{Carmi:2016wjl, Reynolds:2016rvl, Brown:2017jil, Alishahiha:2015rta, Chapman:2016hwi, Chapman:2018dem, Li:2020ark, Engelhardt:2021mju, Frey:2021cjs}.

Motivated by the holographic duality, the physics of quantum circuit complexity from the sides of quantum field theory (QFTs) and quantum mechanics have attracted more and more attentions. For a quantum chaotic system, the elementary physical quantities, like the scrambling time and Lyapunov exponent, can be captured by out-of-time-order correlators (OTOCs) \cite{Maldacena:2015waa,Hayden:2007cs,Sekino:2008he,Lashkari:2011yi,Ali:2019zcj,Bhattacharyya:2020art,Morita:2021syq}. Some recent works, for instance \cite{Bhattacharyya:2020art,Ali:2019zcj}, show that the evolution of quantum circuit complexity could provide equivalent information about the classical scrambling time and Lyapunov exponent in analogous to the OTOCs. There exist three main geometrized methods in calculating complexity for the Gaussian quantum states \cite{Ali:2018fcz}. In areas of quantum information/computation, a type of geometric way is developed by Nielsen and his collaborators to compute the quantum circuit complexity, including the wave-function and the covariance matrix approachs respectively \cite{NielsenComplexity1,NielsenComplexity2,NielsenComplexity3}. Recently, the state complexity in QFTs is defined and quantified through the distance measures in Fubini-Study metric \cite{Chapman:2017rqy,Jefferson:2017sdb}.

The $D_p$ branes play an important role in various aspects of theoretical physics. In string theory, $D_p$ brane is a (p+1)-dimensional extended objects upon which the open strings end up with Dirichlet boundary conditions\cite{Polchinski:1996na}. One of the most amazing properties of $D_p$ brane is that the $U(1)$ Abelian gauge symmetry of a single brane is promoted to be a non-Abelian $U(N)$ symmetry for $N$ coincident branes \cite{Witten:1995im,Myers:1999ps}. Because of this property, the non-perturbative studies of (super-) Yang-Mills gauge field can be always associated to the dynamics of stacked $D_p$ branes \cite{Tseytlin:1998cq, Gubser:1998nz, Kiritsis:1999tx}. In particular, black $p$-brane solutions of type $IIB$ supergravity are constructed by  \cite{Horowitz:1991cd}, which is emerged from the stacks of a number of $D_p$ branes. And the $\mathcal{N}=4$ super-Yang-Mills(SYM) theory in the large $N$ limit could be associated to the type $IIB$ superstring propagating on the near-horizon geometry of black $3$-brane ($AdS_5\times S^5$) \cite{Maldacena:1997re, Gibbons:1993sv,Aharony:1999ti}. From the viewpoints of thermodynamics, the physics of thermal field theory of $\mathcal{N}=4$ SYM theory in the large $N$ limit are consistent with the thermodynamics of AdS-Schwarzchild black holes \cite{Witten:1998zw}. Besides, as the topological defects in spacetime, $D_p$ branes (especially the $D_3$ brane) are frequently used to consider the cosmological phenomenology, for instance the $D_3/\bar{D}_3$ inflation \cite{Kachru:2003sx,Dvali:2001fw,Burgess:2001fx,Alexander:2001ks}, construction of bouncing universe \cite{Kachru:2002kx,Sen:2003mv}. If the Standard Model (SM) gauge bosons are assumed to be originated from the fluctuations of the $D_3$ brane, one could think the SM universe as a probe D-brane moving in the geometry emerged from collection of coincident branes, i.e. the black $3$-brane solution. Meanwhile, the gauge interactions are localized on universe brane and gravity lives in the bulk space. The geodesic motion of universe brane toward outgoing or ingoing directions correspond to the cosmological expansion or contraction respectively. This scenario is known as the Mirage cosmology \cite{Kehagias:1999vr,Papantonopoulos:2000yz,Steer:2002ng,Boehm:2002kf,Germani:2006pf,Germani:2008zh,Germani:2007ub}.

Although the physical picture of Mirage cosmology is analogous to the braneworld cosmology, there has some subtleties \cite{Kehagias:1999vr}. In the model of braneworld cosmology (especially the one basing on $RS$-$II$ model), the movement of univese brane is driven by the 4-dimensional self-gravity localized on brane \cite{Binetruy:1999ut,Csaki:1999jh,Cline:1999ts}. However, the case of Mirage cosmology is different. Since the black $3$-brane geometry is emerged from the stacks of coincident branes, and hence a moving probe $D_3$-brane in black $3$-brane background could be viewed as a elastic collision between the stacked branes at adjacent spacetime positions $X^\mu(\tau_i)$ and $X^\mu(\tau_{i+1})$, which seems like the dominoes. During the elastic collision of branes, besides the transmission of energy, there also has information transfer. In particular, when quantum effects are considered, the information could be encoded into the quantum states localized on the brane. Given a simple reference state at the initial time $\tau_0$, it will evolve into the desired target state at the late time $\tau_f$ through a series of unitary transformations $U(\tau_i\to \tau_{i+1})$. The unitary transformation is generated by the perturbative Hamiltonian which is derived from the quantum fluctuations around the moving $D_3$ brane. In this paper the reference state and target state are chosen as vacuum state and squeezed quantum state respectively, while our purpose is to evaluate the quantum circuit complexity between them by using the wave-function method. The complexity in quantum computation refers to the cost-effective operations necessary to produce the desired target state from a given reference state. More importantly, as a diagnostic for quantum chaos, this physical quantity could also tell us whether the quantum fluctuations around the moving $D_3$ brane will behave as quantum chaotic system at late time. Our results show that the quantum fluctuations around the non-BPS brane will evolve into the chaotic regime unavoidably at the late time, while the chaotic behavior is hard to observe in case of BPS brane.

Our work is structured as follows. In section \ref{reviewBpbrane}, we briefly review the basic $3$-brane solution derived from the bosonic part of the type $IIB$ supergravity. Besides, the corresponding conserved quantities and thermodynamic variables are computated. In section \ref{MovingDbrane}, we first describe the motion of probing $D3$-brane in background of near horizon limit of black $3$-brane geometry. And then the small quantum fluctuations around the classical trajectory of probing $D3$-brane are considered, while the quadratic perturbative action is constructed. The squeezed quantum (SQ) state is introduced in section \ref{FinalComplexity}. The evolution of parameters characterizing the SQ state are governed by the Hamiltonian operator derived from the quadratic perturbative action. And then the complexity between initial vacuum state and the SQ state evolved by the quantum fluctuations is evaluated by using the wave-function approach. The conclusions and discussions are given in the last section.

\section{Bulk spacetime : black $D_3$-brane solution and  conserved quantities \label{reviewBpbrane}} 

In the string frame, a near-extremal black 3-brane in 10-dimensional asymptotically flat spacetime is solved from the equations of bosonic part of the type $\uppercase\expandafter{\romannumeral2} B$ supergravity
\begin{align}
\label{BD3brane}
&ds_{10}^{2}=H_{3}^{-1/2}\big(-Fdt^{2}+d\vec{x}\cdot d\vec{x}\big)+H_{3}^{1/2}\big(\frac{dr^{2}}{F}+r^{2}d\Omega_{5}^{2}\big)\\
\nonumber
\\
\label{BD3braneSol}
&H_{3}(r)=1+\frac{L^{4}}{r^{4}}~,~F(r)=1-\frac{r_{h}^{4}}{r^{4}}\\
\nonumber
\\
\nonumber
&d\Omega_{5}^{2}=d\theta_{1}^{2}+\sin^{2}\theta_{1}(d\theta_{2}^{2}+\sin^{2}\theta_{2}(d\theta_{3}^{2}+\sin^{2}\theta_{3}(d\theta_{4}^{2}+\sin^{2}\theta_{4}d\theta_{5}^{2})))
\end{align}
Actually, this geometry is emerged by a number of coinciding $D_3$-branes. The parameters $L$ and $r_h$ represent the $AdS_5$ curvature radius and the position of horizon respectively. Besides, there also exist the Ramond-Ramond self-dual 5-form field strength $F_{(5)}$ and the dilaton scalar field $\Phi$. As shown by the \cite{Horowitz:1991cd}, the dilaton field could be an arbitrary constant value. At the same time, the $F_{(5)}$ satisfies the following equation
\begin{align}
&\partial_{\mu}\big(\sqrt{-g}g^{\mu\mu_{1}}g^{\alpha\alpha_{1}}g^{\beta\beta_{1}}g^{\gamma\gamma_{1}}g^{\rho\rho_{1}}F_{\mu_{1}\alpha_{1}\beta_{1}\gamma_{1}\rho_{1}}\big)=0
\end{align}
Together with the relation $F_{(5)}=dC_{(4)}$ and the assumption that $C_{(4)}$ is only the function of $r$. The 4-form potential $C_{\mu\nu\rho\sigma}$ is solved as
\begin{align}
\label{SolC4}
&C_{0123}(r)=\sqrt{1+\frac{r_{h}^{4}}{L^{4}}}\frac{H_{3}(r)-1}{H_{3}(r)}
\end{align}
where the constant $\sqrt{1+\frac{r^4_h}{L^4}}$ is derived from the Einstein field equations. And the corresponding field strength $F_{(5)}$ is
\begin{align}
\label{FieldStreF5D3}
&F_{r0123}=-\frac{4L^{2}r^{3}\sqrt{L^{4}+r_{h}^{4}}}{(L^{4}+r^{4})^{2}}
\end{align}
Furthermore, the field strength in the transverse directions is implied by the self-duality condition 
\begin{align}
\label{SelfDualTrans}
&F^{(\Omega_5)}_{\mu_{1}\dots\mu_{5}}=(^{\star}F)_{\mu_{1}\dots\mu_{5}}=\frac{1}{5!\sqrt{-g}}\epsilon_{\mu_{1}\dots\mu_{5}\nu_{1}\dots\nu_{5}}F^{\nu_{1}\dots\nu_{5}}
\end{align}
in which we use the convention $\epsilon^{012\dots}=1$. Substitute $\eqref{BD3brane}$-$\eqref{BD3braneSol}$ and $\eqref{FieldStreF5D3}$ into the $\eqref{SelfDualTrans}$, we give
\begin{align}
\label{DualFieldStreOme5}
&F_{45678}^{(\Omega_{5})}=-\frac{4L^{4}}{5!}\sqrt{1+\frac{r_{h}^{4}}{L^{4}}}\sin^{4}\theta_{1}\sin^{3}\theta_{2}\sin^{2}\theta_{3}\sin\theta_{4}
\end{align}

\begin{figure}[ht]
	\begin{center}
		\includegraphics[scale=0.53]{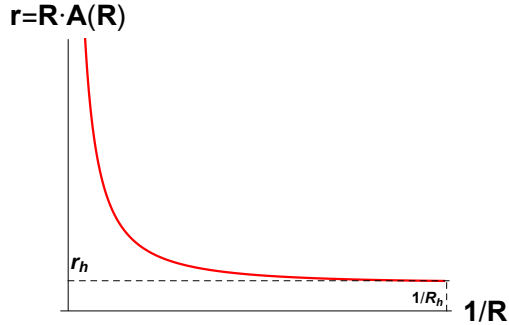}
		\caption{(color online). The variation trend of $r=RA(R)$ with respect to the $\frac{1}{R}$ solved from constraint equation $\eqref{eqsAR}$ together with the boundary conditions $ A(R)_{R\to\infty}=1~ \&~ r_{h}=A(R_h) R_h$.}
		\label{RelarR}
	\end{center}
\end{figure}

The stress-energy tensor for the $3$-brane world volume is given by the \cite{Myers:1999psa} as
\begin{align}
\label{STenergy}
&T_{ab}=\frac{1}{2\kappa^{2}}\int d\Omega_{5}~r^{5}n^{i}\big\{\eta_{ab}(\partial_{i}h_{~\mu}^{\mu}-\partial_{j}h_{~i}^{j})-\partial_{i}h_{ab}\big\}\bigg\vert_{r\to\infty}		
\end{align}
where $h_{\mu\nu}=g_{\mu\nu}-\eta_{\mu\nu}$ represents the deviation of the black $3$-brane metric from that for the flat spacetime, while $n^i$ is a radial unit vector in a direction transverse to the $3$-brane world volume. It is necessary to indicate that the labels $a,b=0,1,2,3$ denote the $3$-brane world volume directions, while $i,j=4,5\dots,9$ run over the transverse directions. Besides, note that the eqs.$\eqref{STenergy}$ is only appliable in asymptotically Cartesian coordinates. For applying the eqs.$\eqref{STenergy}$, we transform the metric $\eqref{BD3brane}$ into the following form,
\begin{align}
&ds_{10}^{2}=H_{3}^{-1/2}(R)\big(-F(R)dt^{2}+d\vec{x}\cdot d\vec{x}\big)+H_{3}^{1/2}(R)A(R)^{2}\delta_{ij}dy^{i}dy^{j}\\
\label{restrictR}
&\frac{dr^{2}}{F(r)}=A(R)^{2}dR^{2}~,~r=A(R)R\\
\nonumber
&y^{1}=R\cos\theta_{1},~y^{2}=R\sin\theta_{1}\cos\theta_{2},~y^{3}=R\sin\theta_{1}\sin\theta_{2}\cos\theta_{3},~y^{4}=R\sin\theta_{1}\sin\theta_{2}\sin\theta_{3}\cos\theta_{4}\\
\label{CardeTran}
&y^{5}=R\sin\theta_{1}\sin\theta_{2}\sin\theta_{3}\sin\theta_{4}\cos\theta_{5}~,~y^{6}=R\sin\theta_{1}\sin\theta_{2}\sin\theta_{3}\sin\theta_{4}\sin\theta_{5}
\end{align}
From the $\eqref{BD3braneSol}$ and $\eqref{restrictR}$, we could derive the differential equation for $A(R)$ as
\begin{align}
\label{eqsAR}
&A^{2}R^{6}A^{\prime2}+2A^{\prime}A^{3}R^{5}=-r_{h}^{4}
\end{align}
Together with the boundary conditions $ A(R)_{R\to\infty}=1~ \&~ r_{h}=A(R_h) R_h$, the relation between $r$ and $R$ is sloved numerically as Fig.\ref{RelarR}. 
Analytically, the above boundary conditions and equation $\eqref{eqsAR}$ imply the following series solution for $A(R)$
\begin{align}
&A(R)=\sum_{n=0}^{\infty}c_{n}(\frac{r_{h}^{4}}{R^{4}})^{n},~c_{1}=\frac{1}{8c_{0}^{3}},~c_{2}=-\frac{1}{128c_{0}^{7}},~c_{3}=\frac{1}{1024c_{0}^{11}},~\dots
\end{align}
By rescaling the $R$, we could set the value of dimensionless parameter $c_0$ to be $1$. From the coordinate transformations $\eqref{CardeTran}$, the radial unit vector in transverse direction is constructed as $n^i=\frac{y^i}{R}$. Meanwhile $\partial_i$ is calculated as $\partial_i=\frac{\partial R}{\partial y^i}\cdot\frac{\partial}{\partial R}=\frac{y^i}{R}\frac{\partial}{\partial R}$ by differentiating the equality $R^2=\sum_{i} y^i\cdot y^i$. The ADM mass $M$ of black 3-brane is associated to the $00$ component of $T_{ab}$ through the ralation
\begin{align}
 &M=V_3 T_{00}=\frac{\Omega_{5} V_3 }{2\kappa^{2}}\big\{-\eta_{00}(2L^{4}+r_{h}^{4})+2L^{4}+4r_{h}^{4}\big\}
 \end{align}
 in which $V_3$ and $\Omega_5$ represent the volume of 3-dimensional D-brane and the volume of a unit 5-dimensional sphere $S^5$ respectively. The Ramond-Ramond charge $N$ (integer) which is the source of the form field $C_{(4)}$ could be evaluated through the Gauss's law
 \begin{align}
 \label{GaussLawCacRRcharge}
 &\lim_{r\to \infty}\int_{\Omega^{5}}dx^{\nu_{4}}\wedge\dots\wedge dx^{\nu_{8}}F_{\nu_{4}...\nu_{8}}^{(\Omega_{5})}=-2\kappa_{10}^{2}T_{3}N
 \end{align}
Plugging the equation $\eqref{DualFieldStreOme5}$ into the $\eqref{GaussLawCacRRcharge}$, the charge $N$ is obtained as
\begin{align}
&N=\frac{2L^{4}\Omega^{5}}{\kappa_{10}^{2}T_{3}}\sqrt{1+\frac{r_{h}^{4}}{L^{4}}}
\end{align}
According to standard methods of black hole thermodynamics, the Hawking temperature and entropy are calculated as
\begin{align}
&T=\frac{(H_{3}^{-1/2}F)^{\prime}}{4\pi}\bigg\vert_{r=r_{h}}=\frac{r_{h}}{\pi\sqrt{r_{h}^{4}+L^{4}}}\\
&S=\frac{A}{4G}=\frac{1}{4G}\int d\Omega_{5}dV_{3}r_{h}^{5}H_{3}^{1/2}(r_{h})=\frac{2\pi V_{3}\Omega_{5}}{\kappa^{2}}r_{h}^{3}\sqrt{r_{h}^{4}+L^{4}}
\end{align}
Besides, the chemical potential of $C_{(4)}$ is written as
\begin{align}
&\Phi=-V_{3}T_{3}\big(C_{0123}(r)\big\vert_{r\to\infty}-C_{0123}(r)\big\vert_{r\to r_{h}}\big)=\frac{V_{3}T_{3}L^{2}}{\sqrt{L^{4}+r_{h}^{4}}}
\end{align}
Finally, one could show that these thermodynamics quantities indeed satisfy the first law of black hole thermodynamics
\begin{align}
&dM=TdS+\Phi dN
\end{align}

In the near horizon limit $TL\ll 1$, the metric $\eqref{BD3brane}$ describes $AdS_5 \times S^5$ geometry
\begin{align}
\label{NearHorizonBD3}
&ds_{10}^{2}=\frac{r^{2}}{L^{2}}\big(-(1-r_{h}^{4}/r^{4})dt^{2}+d\vec{x}\cdot d\vec{x}\big)+\frac{L^{2}}{r^{2}}\frac{dr^{2}}{(1-r_{h}^{4}/r^{4})}+L^{2}d\Omega_{5}^{2}		
\end{align}
From the viewpoint of holography \cite{Witten:1998qj}, the geometry $\eqref{NearHorizonBD3}$ is dual to the boundary conformal field theory related to the large N dynamics of broken $\mathcal{N}=4$ Supersymmetric-Yang-Mills (SYM) theories at non-zero temperature. Meanwhile, the limiting value of the 4-form potential $\eqref{SolC4}$ is
\begin{align}
\label{NearHorizonC0123}
&C_{0123}(r)=1-\frac{r^{4}}{L^{4}}+\frac{1}{2}\frac{r_{h}^{4}}{L^{4}}+O(\frac{r^{8}}{L^{8}})
\end{align} 
In scenario of AdS/SYM correspondence, a $D_3$-brane probe in the bulk of the AdS spacetime corresponds to the spontaneous symmetry breaking from $U(N+1)$ to $U(N)\times U(1)$ triggered by the vacuum expectation value of Higgs.

\section{The motion of probe $D_3$ brane \label{MovingDbrane}}

\subsection{classical dynamics and Friedmann equation}

In this part, we consider a probing D3-brane moving in the background spacetime $\eqref{NearHorizonBD3}$. For simplification, the probe is assumed to move only along the radial direction $r$ and have no dynamics around $S^5$. Thus in the rest of this paper we just consider the bulk spacetime with metric
\begin{align}
&ds_5 ^2=g_{\mu\nu}dX^\mu dX^\nu=-f(r)dt^2+\chi(r)d\vec{x}\cdot d\vec{x}+h(r)dr^2
\end{align}
where
\begin{align}
\label{SolNearHorizonD3}
&f(r)=\frac{r^{2}}{L^{2}}(1-\frac{r_{h}^{4}}{r^{4}})\quad,\chi(r)=\frac{r^{2}}{L^{2}}\quad,h(r)=\frac{1}{f(r)}
\end{align}
We use $X^\mu=\{t,\vec{x},r\}=\{x^a,r\}$ and $\sigma^a=\{t_\sigma,\vec{\sigma}\}$ to denote bulk coordinates and intrinsic worldsheet coordinates respectively. Here, we consider an infinitely straight probing brane parallel to the $x^a$ hyperplane but free to move along the radial direction $r$. And hence the embedding way is
\begin{align}
\label{EmbedDbrane}
& t=t_\sigma \quad,\quad \vec{x}=\vec{\sigma}\quad,\quad r=R(t)
\end{align}
The induced metric on the brane is
\begin{align}
\label{cacInduceMet}
&\gamma_{ab}=g_{\mu\nu}(X)\frac{\partial X^{\mu}}{\partial\sigma^{a}}\frac{\partial X^{\nu}}{\partial\sigma^{b}}
\end{align}
and then the brane worldsheet is
\begin{align}
\label{InduMetricOnBrane}
&ds^2_4=\gamma_{ab}dx^a dx^b=-\big(f(R)-h(R)\dot{R}^{2}\big)dt^{2}+\chi(R)d\vec{x}\cdot d\vec{x}
\end{align}
Actually, the $\eqref{InduMetricOnBrane}$ could be transformed to the FLRW metric via the redefinition
\begin{align}
\label{DefineDtau}
&ds^{2}=-d\tau^{2}+a(\tau)^{2}d\vec{x}\cdot d\vec{x} \quad \quad d\tau=\sqrt{f(R)-h(R)\dot{R}^{2}}dt~,~a(\tau)=\sqrt{\chi(R(\tau))}
\end{align}
For the convenience of calculations in later, we give a useful relation
\begin{align}
\label{dRdtAnddRdtau}
&\dot{R}^{2}=\frac{f}{h+\frac{1}{R_{,\tau}^{2}}}
\end{align}
in which we use $\dot{R}$ and $R_{,\tau}$ to denote $\frac{dR}{dt}$ and $\frac{dR}{d\tau}$ respectively.

\begin{figure}[ht]
	\begin{center}
		\includegraphics[scale=0.53]{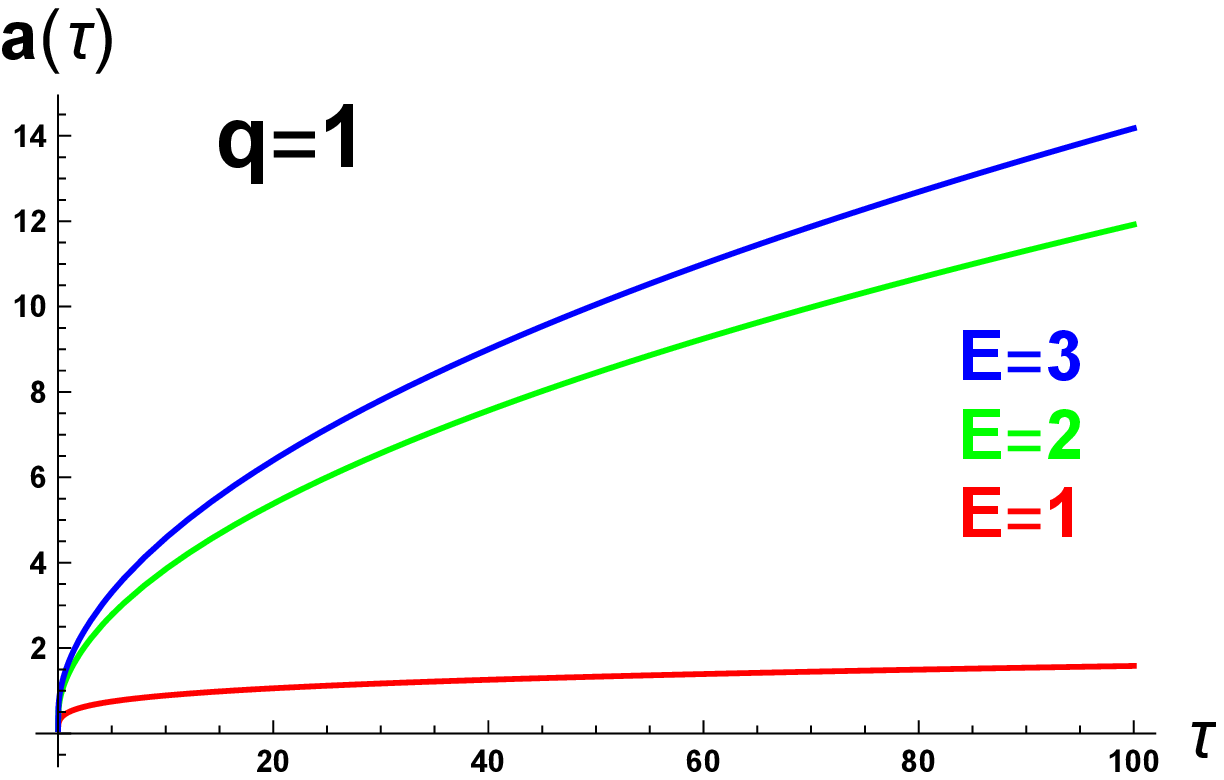}
		\includegraphics[scale=0.53]{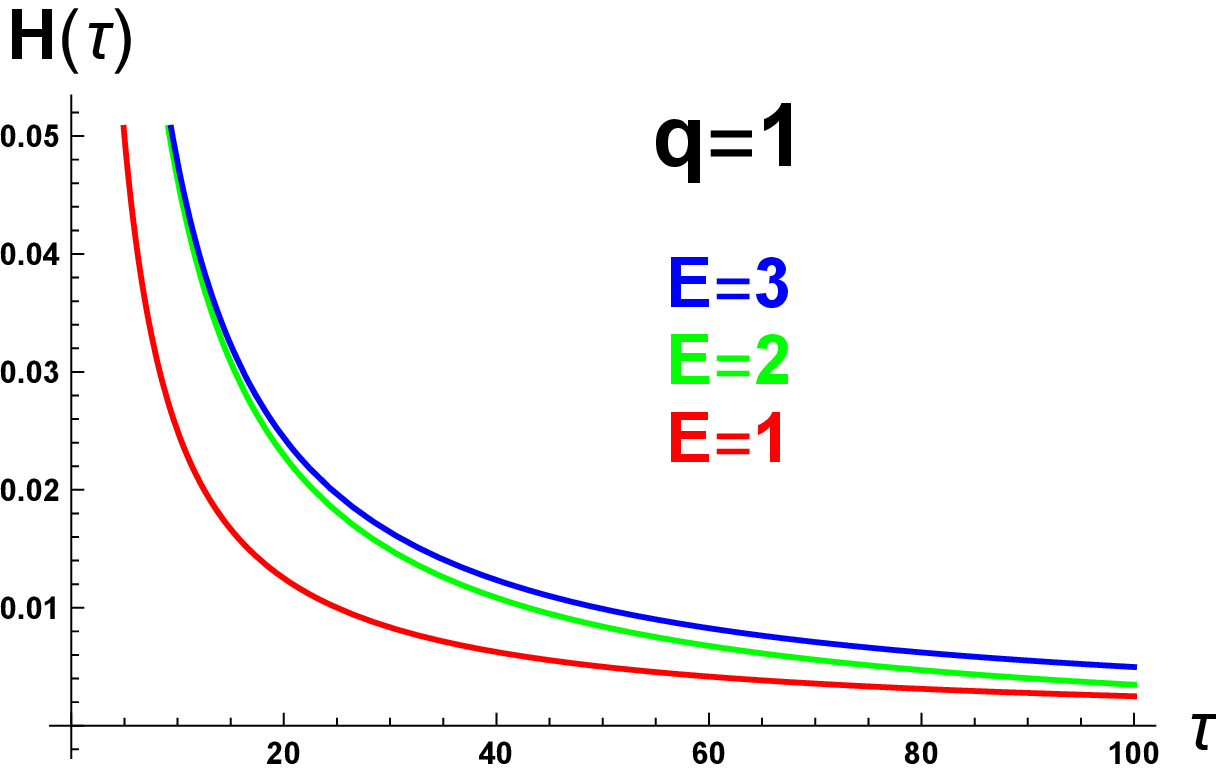}\\
		\includegraphics[scale=0.53]{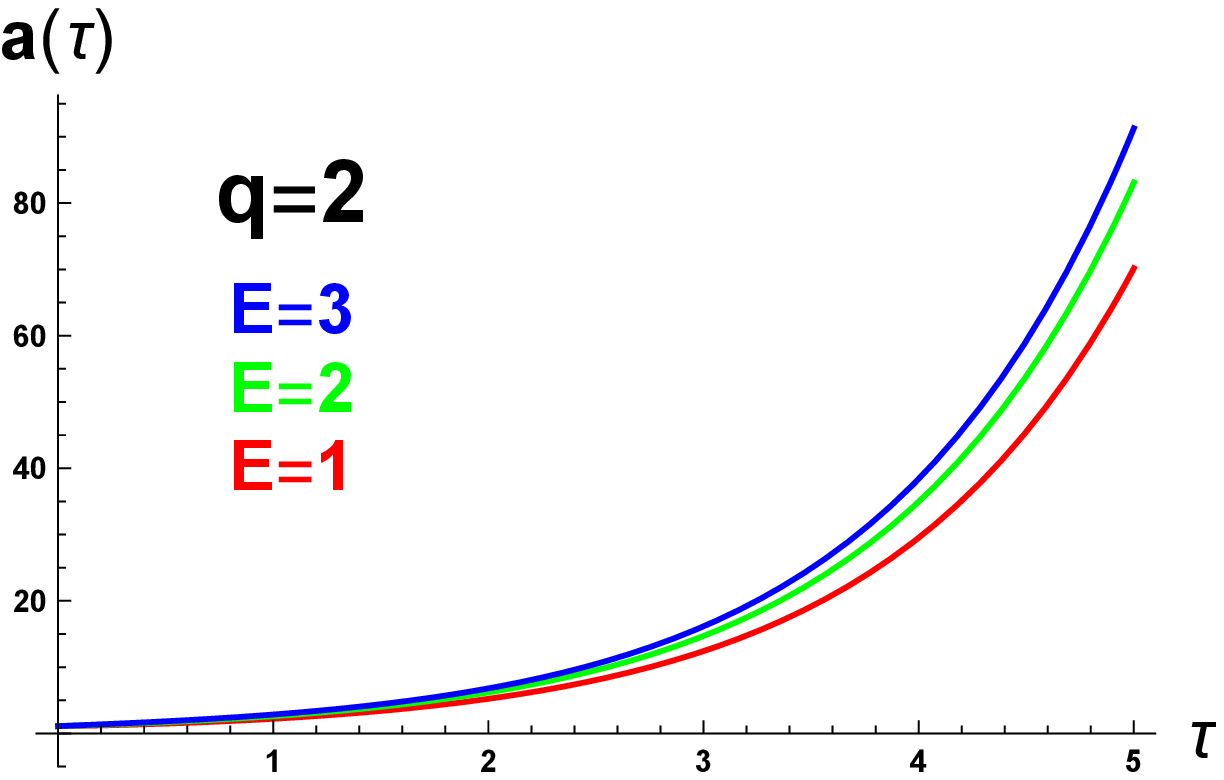}
		\includegraphics[scale=0.53]{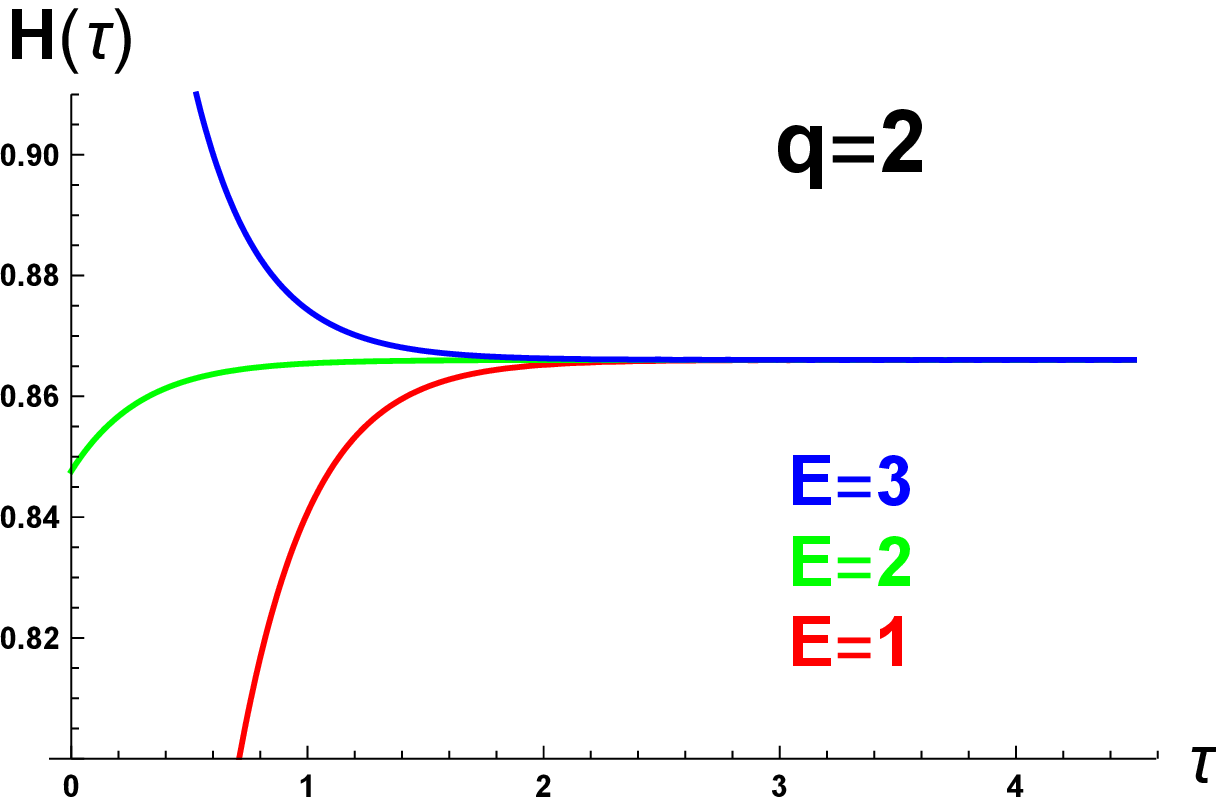}
		\caption{(color online). Numerical solutions of a Friedmann-like equation $\eqref{FriedMannForD3}$ for $r_h=1$, $L=2$, $q=1,2$ and some representative values of $E$.}
		\label{SolFriedInBPS}
	\end{center}
\end{figure}

The standard Dirac-Born-Infeld (DBI) action for type IIB superstring theory is
\begin{align}
\label{ComDBI}
&S_{probeD_{3}}=-T_{3}\int d^{4}\sigma\sqrt{-\det\big(\gamma_{ab}+2\pi\alpha^{\prime}F_{ab}+B_{ab}\big)}-\rho_{3}\int C_{(4)}
\end{align}
in which $F_{ab}$ is the field strength tensor of the gauge fields on the brane, while the quantities $\hat{B}_{ab}$ and $\hat{C}_{4}$ represent the pull-back of the Neveu-Schwatz (NS) 2-form field and the RR 4-form potential in the bulk. Besides, $\rho_3=qT_3$ is the brane charge under a RR 4-form potential living in bulk, in which $q=\pm 1$ correspond to the BPS branes and BPS anti-branes respectively. However, the NS field $B_{\mu\nu}$ vanishes in background geometry $\eqref{BD3brane}$. For the sake of simplification, we neglect the effects of gauge field $F_{ab}$ on the brane. So the DBI action $\eqref{ComDBI}$ reduces to
\begin{align}
\label{ProbeD3Action}
&S_{probeD_{3}}=T_{3}\int d^{4}\sigma\mathcal{L}_{probeD_{3}}=-T_{3}\int d^{4}\sigma\sqrt{-\det\gamma_{ab}}-\rho_{3}\int C_{(4)}
\end{align}
where
\begin{align}
\label{GeneDefineC4}
&C_{(4)}=\frac{1}{4!}C_{\mu\nu\alpha\beta}\frac{\partial X^{\mu}}{\partial\sigma^{a}}\frac{\partial X^{\nu}}{\partial\sigma^{b}}\frac{\partial X^{\alpha}}{\partial\sigma^{c}}\frac{\partial X^{\beta}}{\partial\sigma^{e}}d\sigma^{a}\wedge d\sigma^{b}\wedge d\sigma^{c}\wedge d\sigma^{e}
\end{align}
Substitute $\eqref{NearHorizonC0123}, \eqref{EmbedDbrane}, \eqref{GeneDefineC4}$ into $\eqref{ProbeD3Action}$, the Lagrangian $\mathcal{L}_{probeD_{3}}$ reads
\begin{align}
\label{LagranProbe}
&\mathcal{L}_{probeD_{3}}=-\sqrt{\big(f(R)-h(R)\dot{R}^{2}\big)\cdot g^{3}(R)}+q\frac{R^{4}}{L^{4}}-q(1+\frac{r_{h}^{4}}{2L^{4}})
\end{align}
Since the $\eqref{LagranProbe}$ does not depend on time explicitly, a conserved Hamiltonian density is constructed as
\begin{align}
\label{ConHamiltonian}
&\mathcal{H}=\frac{\partial\mathcal{L}}{\partial\dot{R}}\dot{R}-\mathcal{L}=q(1+\frac{r_{h}^{4}}{2L^{4}})-q\frac{R^{4}}{L^{4}}+\frac{f\sqrt{\chi^{3}}}{\sqrt{f-h\dot{R}^{2}}}=E
\end{align}
From the $\eqref{ConHamiltonian}$, it is straightforward to calculate
\begin{align}
\label{D3BraneMovesV1}
&\dot{R}^{2}=\frac{f}{h}\bigg(1-\frac{f\chi^{3}}{\big(E-q(1+\frac{r_{h}^{4}}{2L^{4}})+q\frac{R^{4}}{L^{4}}\big)^{2}}\bigg)
\end{align}
Transforming to the proper time of brane via $\dot{R}^{2}=\frac{f}{h+\frac{1}{R_{\tau}^{2}}}$, the above equation yields
\begin{align}
\label{D3BraneMovesV2}
&R_{\tau}^{2}=\frac{\big(E-q(1+\frac{r_{h}^{4}}{2L^{4}})+q\frac{R^{4}}{L^{4}}\big)^{2}}{hf\chi^{3}}-\frac{1}{h}
\end{align}
where $R_{\tau}$ denotes the $\frac{dR}{d\tau}$. Combine $\eqref{SolNearHorizonD3}$ with $\eqref{D3BraneMovesV2}$, a Friedmann-like equation is derived as \cite{Kehagias:1999vr}
\begin{align}
\label{FriedMannForD3}
&H^{2}=\frac{1}{L^{2}}\bigg(\frac{\tilde{E}^{2}}{a^{8}}+\frac{2q\tilde{E}+\frac{r_{h}^{4}}{L^{4}}}{a^{4}}+(q^{2}-1)\bigg)	
\end{align}
in which $a(\tau)=\frac{R(\tau)}{L}$ and $\tilde{E}=E-\frac{qr^4_h}{2L^4}-q$. In $\eqref{FriedMannForD3}$, the term in $1/a^8$ is named by dark fluid state, which dominates the evolution at early times \cite{Steer:2002ng}. In analogy to the brane cosmology with $Z_2$ symmetry, the part including $r_h^4$ in second term is associated with the projection operation of bulk Weyl tensor \cite{Kraus:1999it}. Meanwhile, the part proportial to $\tilde{E}$ is originated from the broken of $Z_2$-symmetry \cite{Steer:2002ng}. The last term in $\eqref{FriedMannForD3}$ represents an effective 4-dimensional cosmological constant, which will vanish when D brane is in BPS state (namely $q=\pm1$). The numerical solutions corresponding to $\eqref{FriedMannForD3}$ in some representative $E$ parameter with $q=1,2$ separately are shown by Fig.\ref{SolFriedInBPS}. It is easy to see that both BPS brane and non-BPS brane will move outward to the infinity, but with decreased velocity and increased velocity respectively.
Note that the anti-branes ($q\leq-1$) will move inward to the horizon and falls into the black brane at the end.

\subsection{Perturbative action}

\begin{figure}[ht]
	\begin{center}
		\includegraphics[scale=0.53]{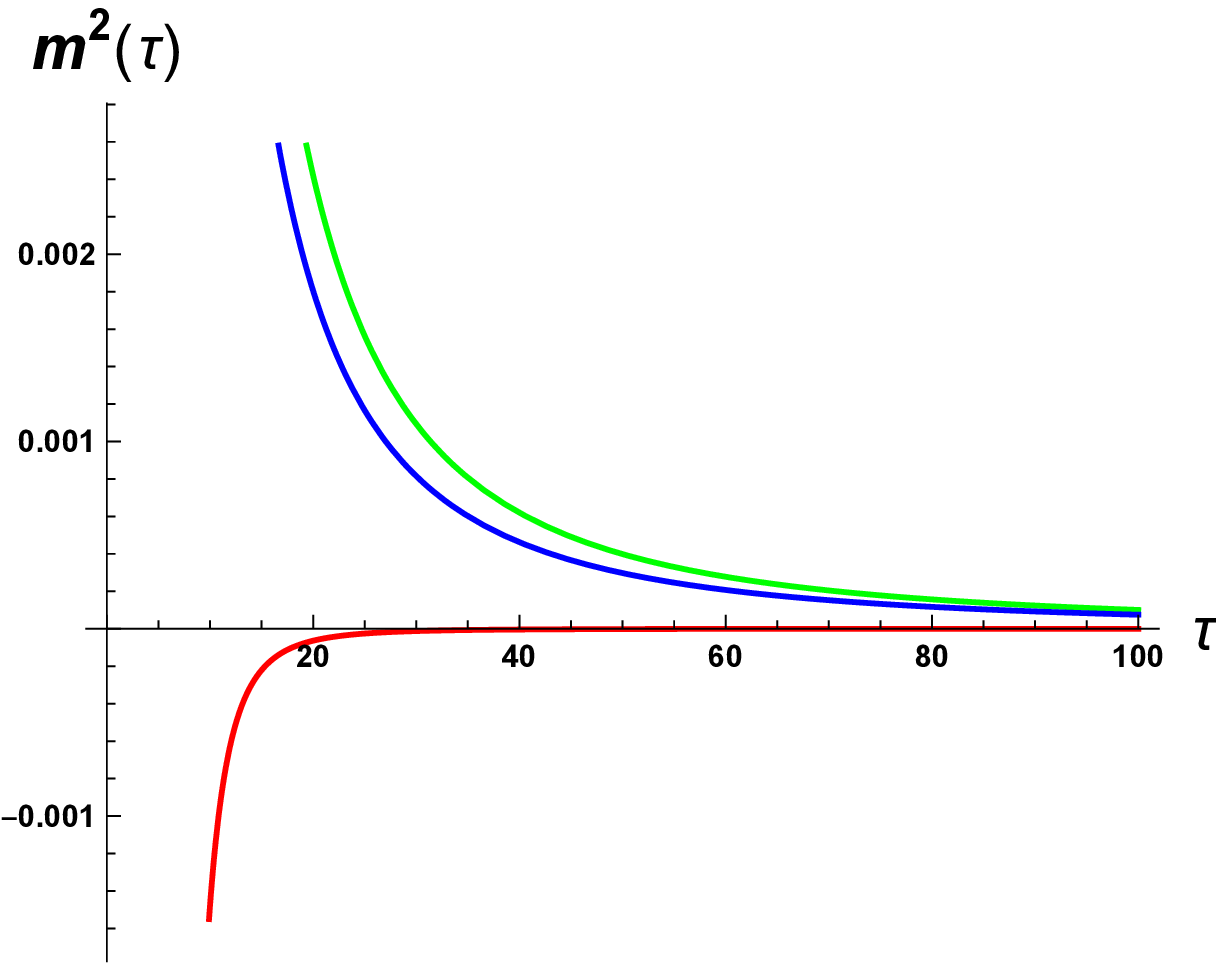}
		\includegraphics[scale=0.53]{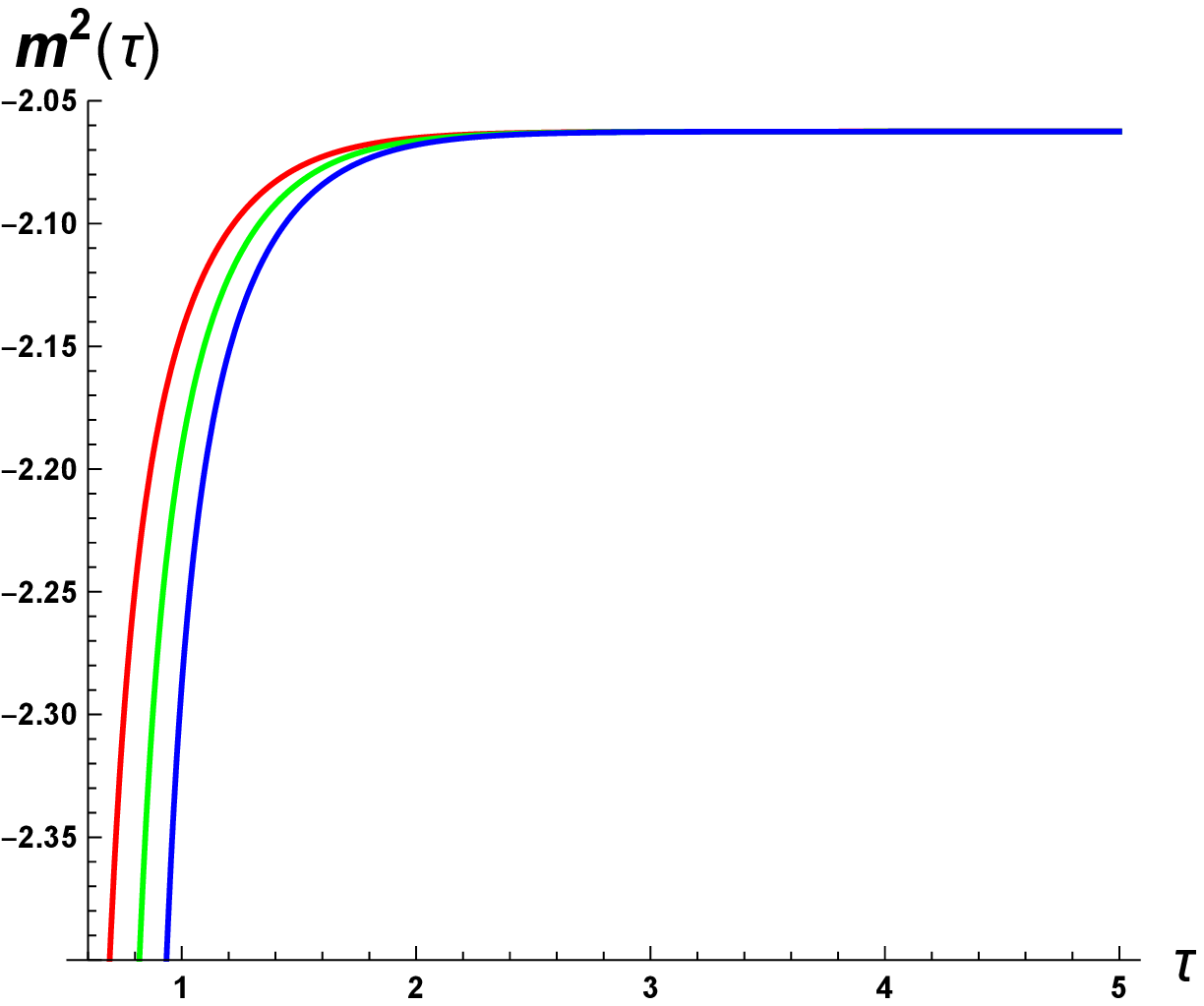}
		\caption{(color online). The effective mass square $m^2(\tau)$ as a function of brane proper time $\tau$, which are consistent with the parameters and solutions given in Fig.\ref{SolFriedInBPS}. The left panel and right panel correspond to the $q=+1$ and $q=2$ respectively.}
		\label{EffecSquMass}
	\end{center}
\end{figure}

Here we will use the symbols with bar to denote the unperturbed classical quantities, while those without bar representing the perturbed ones. In term of embedding coordinates in the bulk spacetime, fluctuations around the moving probe $D_{3}$ brane is given by
\begin{align}
\label{FluctNearBulkEmbed}
&X^{\mu}(t,\vec{x})=\bar{X}^{\mu}(t)+\Phi(t,\vec{x})\bar{n}^{\mu}(t)
\end{align}
where $\bar{n}^{\mu}$ is the unit normal into the motional $D_3$ brane. The velocity of the brane in term of the proper time $\tau$ is written as
\begin{align}
&\bar{\mu}^{\alpha}=\big\{\dot{t}(\tau),0,0,0,\dot{R}(\tau)\big\}
\end{align}
The normalization $\bar{u}^{\alpha}\bar{u}_{\alpha}=-1$ implies
\begin{align}
\label{pdtpdtau}
&\dot{t}(\tau)=\sqrt{\frac{1+h(R)\dot{R}(\tau)^{2}}{f(R)}}
\end{align}
Actually, the equation $\eqref{pdtpdtau}$ could also be acquired by combining $\eqref{DefineDtau}$ with $\eqref{dRdtAnddRdtau}$. In the light of the orthogonal and normalized conditions $\bar{u}^{\mu}\bar{n}_{\mu}=0~\&~\bar{n}^{\mu}\bar{n}_{\mu}=1$, it reads
\begin{align}
\label{SolNormal}
&\bar{n}_{\mu}=\big\{\sqrt{f(R)h(R)}\dot{R}(\tau),0,0,0,-\sqrt{h(R)\big(1+h(R)\dot{R}(\tau)^{2}\big)}\big\}		
\end{align}
Besides, the projection tensor is written as
\begin{align}
&\bar{\gamma}_{\mu\nu}=\bar{g}_{\mu\nu}- \bar{n}_\mu \bar{n}_\nu
\end{align}
which corresponds to the induced metric $\bar{\gamma}_{ab}$ expressed in term of the indices of bulk spacetime. By combining the $\eqref{cacInduceMet}$ with $\eqref{FluctNearBulkEmbed}$, we obtain the perturbative induced metric up to second order
\begin{align}
\label{PerturInduMetOrder2}
&\gamma_{ab}=\bar{\gamma}_{ab}+\gamma_{ab}^{(1)}+\gamma_{ab}^{(2)}\quad,\quad \gamma_{ab}^{(1)}=2\bar{K}_{ab}\Phi ~,~ \gamma_{ab}^{(2)}=-\Phi_{,a}\Phi_{,b}+\bar{K}_{ab}^{\prime}\Phi^{2}
\end{align}
where the extrinsic curvature tensor $\bar{K}_{ab}$ is defined as
\begin{align}
\label{ExTenBoun}
&\bar{K}_{ab}=\frac{\partial \bar{X}^\mu}{\partial \sigma^a} \frac{\partial \bar{X}^\nu}{\partial \sigma^b}D_{(\mu} \bar{n}_{\nu)} \quad,\quad D_\mu =\bar{\gamma}^\nu _{~~\mu}\nabla_\nu
\end{align}
And $\bar{K}^\prime_{ab}$ represents the Lie derivative $\mathcal{L}_{\bar{n}} \bar{K}_{ab}$. Under the perturbations $\eqref{PerturInduMetOrder2}$, the corresponding  determinant $\det ({\gamma_{ab}})$ is expanded as
\begin{align}
\label{PerturInduDeterMetOrder2}
&\frac{\det(\bar{\gamma}_{ab}+\gamma_{ab}^{(1)}+\gamma_{ab}^{(2)})}{\det(\bar{\gamma}_{ab})}=1+\bar{\gamma}^{ab}\gamma_{ab}^{(1)}+\bar{\gamma}^{ab}\gamma_{ab}^{(2)}+\frac{1}{2}(\bar{\gamma}^{ab}\gamma_{ab}^{(1)})^{2}-\frac{1}{2}\gamma^{(1)ab}\gamma_{ab}^{(1)}+O(\epsilon^{3})
\end{align}
Basing on $\eqref{PerturInduMetOrder2}$ and $\eqref{PerturInduDeterMetOrder2}$, the Nambu-Goto term in $\eqref{ProbeD3Action}$ has the following second order expansions
\begin{align}
\label{GenePerAnsatzV0}
&S^{(2)}=-T_{3}\int d^{4}\sigma\sqrt{-\bar{\gamma}}\big\{\frac{1}{2}\bar{\gamma}^{ab}\Phi_{,a}\Phi_{,b}-\frac{1}{2}\bar{K}^{\prime}\Phi^{2}+\bar{K}^{2}\Phi^{2}\big\}
\end{align}
in which we have used the fact $\bar{K}^{\prime}=\bar{\gamma}^{ab}\bar{K}_{ab}^{\prime}+\bar{K}^{ab}\bar{\gamma}_{ab}^{\prime}$ and $\bar{K}_{ab}=\frac{1}{2}\bar{\gamma}_{ab}^{\prime}=\frac{1}{2}\mathcal{L}_{\bar{n}}\bar{\gamma}_{ab}$. As indicated by work \cite{Guven:1993ew}, together with the background equation of motion, the $\bar{K}^{2}$ term in $\eqref{GenePerAnsatzV0}$ is eliminated by the contribution from perturbations of 4-form field term in $\eqref{ProbeD3Action}$. Besides, basing on the $Gauss$-$Codacci$ equations \cite{Wald:1984}, the $\bar{K}^{\prime}$ could be rewritten as \cite{Guven:1993ew}
\begin{align}
&\bar{K}^{\prime}=D_{\mu}\bar{n}_{\nu}D^{\nu}\bar{n}^{\mu}+\bar{n}^{\nu}D_{\mu}D_{\nu}\bar{n}^{\mu}-\bar{n}^{\mu}D_{\mu}D_{\nu}\bar{n}^{\nu}=\bar{K}^{\mu\nu}\bar{K}_{\mu\nu}+{}^{N}\bar{R}_{\mu\nu}\bar{n}^{\mu}\bar{n}^{\nu}
\end{align}
in which we have used the fact $\bar{K}^{ab}\bar{K}_{ab}=\bar{K}^{\mu\nu}\bar{K}_{\mu\nu}$ and $\bar{\gamma}^{ab}\bar{K}_{ab}^{\prime}=\bar{\gamma}^{\mu\nu}\bar{K}_{\mu\nu}^{\prime}$. Finally, for the action of probe $D_3$ brane, the quadratic perturbative quantity is  
\begin{align}
\label{GenePerAnsatzFin}
&S_{\phi^{2}}=-\frac{1}{2}\int d^{4}x\sqrt{-\bar{\gamma}}\big[\bar{\gamma}^{ab}\partial_{a}\phi\partial_{b}\phi-(\bar{K}^{ab}\bar{K}_{ab}+\bar{R}_{\mu\nu}\bar{n}^{\mu}\bar{n}^{\nu})\phi^{2}\big]
\end{align}
From the definition $\eqref{ExTenBoun}$, the nonvanishing components of $\bar{K}_{ab}$ read
\begin{align}
\label{CacKtautauv1}
&\bar{K}_{\tau\tau}=\frac{f^{\prime}\big(2h^{5/2}f^{1/2}R_{\tau}^{4}+h^{3/2}f^{1/2}R_{\tau}^{2}-f^{1/2}h^{1/2}\big)-h^{\prime}R_{\tau}^{2}\big(h^{1/2}f^{3/2}+2h^{3/2}f^{3/2}R_{\tau}^{2}\big)}{2hf^{3/2}\sqrt{1+hR_{\tau}^{2}}}\\
\label{CacKijv1}
&\bar{K}_{ij}=\frac{1}{2}\sqrt{h^{-1}+R_{\tau}^{2}}\chi^{\prime}\delta_{ij}
\end{align}
After the tedious calculations, the Ricci term is given by
\begin{align}
\nonumber
&\bar{R}_{\mu\nu}\bar{n}^{\mu}\bar{n}^{\nu}=\frac{h^{\prime}f^{\prime}f+hf^{\prime2}-2hff^{\prime\prime}}{4h^{2}f^{2}}+\frac{3(f\chi^{\prime2}+f^{\prime}\chi^{\prime}\chi-2f\chi^{\prime\prime}\chi)R_{\tau}^{2}}{4f\chi^{2}}\\
\label{RicciTerm}
&\quad\quad\quad\quad+\frac{3h^{\prime}\chi^{\prime}\chi+3hh^{\prime}\chi\chi^{\prime}R_{\tau}^{2}+3h\chi^{\prime2}-6h\chi\chi^{\prime\prime}}{4h^{2}\chi^{2}}	
\end{align}
Substituting $\eqref{SolNearHorizonD3}$ into the $\eqref{CacKtautauv1}$-$\eqref{RicciTerm}$ and collecting them together, it yields
\begin{align}
\nonumber
&m^{2}(\tau)=-\bar{K}^{ab}\bar{K}_{ab}-\bar{R}_{\mu\nu}\bar{n}^{\mu}\bar{n}^{\nu}\\
&\quad\quad~~=\frac{4}{L^{2}}-\frac{3(\tilde{E}+a^{4}q)^{2}}{a^{8}L^{2}}-\frac{(\tilde{B}+2a^{4}L^{4})^{2}\big(a^{8}\tilde{B}^{2}+6\tilde{B}\tilde{A}^{2}a^{4}L^{4}+4\tilde{A}^{4}L^{8}\big)^{2}}{\tilde{A}^{2}\tilde{B}^{4}a^{16}L^{10}}
\end{align}
where $a(\tau)=\frac{R(\tau)}{L}$, $\tilde{A}=\tilde{E}+a^{4}$ and $\tilde{B}=r_{h}^{4}-a^{4}L^{4}$, while the $R_\tau^2$ is replaced by the equation $\eqref{D3BraneMovesV2}$. Comparing to the usual quadratic action of cosmological perturbations, there exists an extra effective mass square term in $\eqref{GenePerAnsatzFin}$, which is originated from the projection operation of bulk geometry. As shown by Fig.\ref{EffecSquMass}, the $m^2(\tau)$ could have a positive value only for the BPS-brane.

\section{Complexity for quadratic perturbative action of moving $D_3$ brane \label{FinalComplexity}}

\subsection{The squeezed quantum states}

Transform the brane proper time $\tau$  into the conformal time $\eta=\int d\tau/ a(\tau)$, the perturbative action $\eqref{GenePerAnsatzFin}$ reads
\begin{align}
\label{PerActCon}
&S_{per}=\int d\eta L=\frac{1}{2}\int d^{3}xd\eta\cdot\big[v^{\prime2}-\partial_{i}v\partial_{i}v+(\frac{a^{\prime}}{a})^{2}v^{2}-2\frac{a^{\prime}}{a}v^{\prime}v-m^{2}(\eta)v^{2}\big]
\end{align}
where the prime denotes the derivative respect to the $\eta$. The canonical momentum is obtained from $\eqref{PerActCon}$
\begin{align}
&\pi(\eta,\vec{x})=\frac{\delta L}{\delta v^{\prime}(\eta,\vec{x})}=v^{\prime}-\frac{a^{\prime}}{a}v
\end{align}
So the corresponding Hamiltonian is constructed as
\begin{align}
&H=\int d^{3}x\big[\pi v^{\prime}-\mathcal{L}\big]=\frac{1}{2}\int d^{3}x\big[\pi^{2}+\partial_{i}v\partial_{i}v+\frac{a^{\prime}}{a}(\pi v+v\pi)+m^{2}v^{2}\big]
\end{align}
Promoting the field $v(\eta,\vec{x})$ and $\pi(\eta,\vec{x})$ into operators $\hat{v}(\eta,\vec{x})$, $\hat{\pi}(\eta,\vec{x})$, which obeys the uncertainty relation
\begin{align}
& [\hat{v}(\eta,\vec{x}),\hat{\pi}(\eta,\vec{y})]\big\vert_{\eta=\eta_{0}}=i\delta^{3}(\vec{x}-\vec{y})
\end{align}
The $\hat{v}(\eta,\vec{x})$ and $\hat{\pi}(\eta,\vec{x})$ are supposed to have the following decomposition in Fourier space
\begin{align}
\label{QuanfivField}
&\hat{v}(\eta,\vec{x})=\int\frac{d^{3}k}{(2\pi)^{3/2}}\frac{1}{\sqrt{2k}}\big(\hat{c}_{-\vec{k}}^{\dagger}v_{k}^{\star}(\eta)+\hat{c}_{\vec{k}}v_{k}(\eta)\big)e^{i\vec{k}\cdot\vec{x}}\\
\label{QuanfiEtaField}
&\hat{\pi}(\eta,\vec{x})=i\int\frac{d^{3}k}{(2\pi)^{3/2}}\sqrt{\frac{k}{2}}\big(\hat{c}_{-\vec{k}}^{\dagger}u_{k}^{\star}(\eta)-\hat{c}_{\vec{k}}u_{k}(\eta)\big)e^{i\vec{k}\cdot\vec{x}}
\end{align}
where $k$ denotes the modulus $\vert \vec{k} \vert$, while the creation operator $\hat{c}^\dagger_{-\vec{k}}$ and  annihilation operator $\hat{c}_{\vec{k}}$ satisfy
\begin{align}
&[c_{\vec{p}},\hat{c}_{\vec{q}}^{\dagger}]=(2\pi)^{3}\delta^{3}(\vec{p}-\vec{q})~,~[c_{\vec{p}},c_{\vec{q}}]=0~,~[c_{\vec{p}}^{\dagger},c_{\vec{q}}^{\dagger}]=0
\end{align}
It is necessary to notice that the $\eqref{QuanfivField}$ and $\eqref{QuanfiEtaField}$ are generalization of the inverted harmonic oscillator in quantum mechanics \cite{Barton:1984ey}. By choosing an appropriate normalization condition for mode functions $u_k(\eta)$, $v_k(\eta)$, the Hamiltonian operator is expressed as
\begin{align}
\nonumber
&\hat{H}=\int d^3k \hat{\mathcal{H}}_k=\frac{1}{2}\int d^{3}k\big[\big(k+\frac{m^{2}(\eta)}{2k}\big)(\hat{c}_{\vec{k}}\hat{c}_{\vec{k}}^{\dagger}+\hat{c}_{-\vec{k}}^{\dagger}\hat{c}_{-\vec{k}})\\
\label{SqueeHamiltonian}
&\quad+i\frac{a^{\prime}}{a}(\hat{c}_{-\vec{k}}^{\dagger}\hat{c}_{\vec{k}}^{\dagger}-\hat{c}_{-\vec{k}}\hat{c}_{\vec{k}})+\frac{m^{2}(\eta)}{2k}(\hat{c}_{-\vec{k}}^{\dagger}\hat{c}_{\vec{k}}^{\dagger}+\hat{c}_{-\vec{k}}\hat{c}_{\vec{k}})\big]
\end{align}
The first term in $\eqref{SqueeHamiltonian}$ corresponds to the Hamiltonian of free particle with effective mass, while the second and third terms represent the interaction between the quantum fluctuations and the background of moving $D_3$ brane. From the interaction Hamiltonian operator, it is straightforward to see the momentum structure of the creation and annihilation operators give rise to the production or destruction of particle in pairs with opposite momenta. 

\begin{figure}[ht]
	\begin{center}
		\includegraphics[scale=0.45]{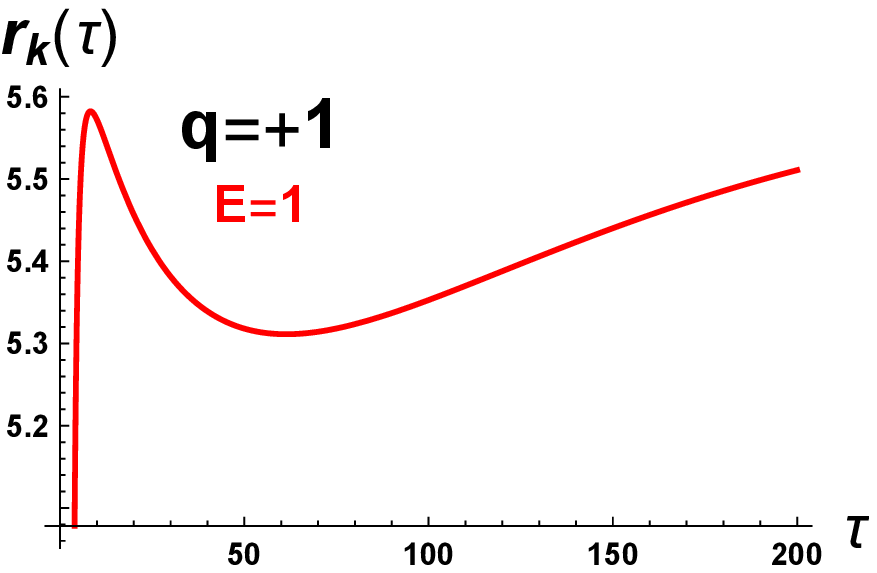}
		\includegraphics[scale=0.45]{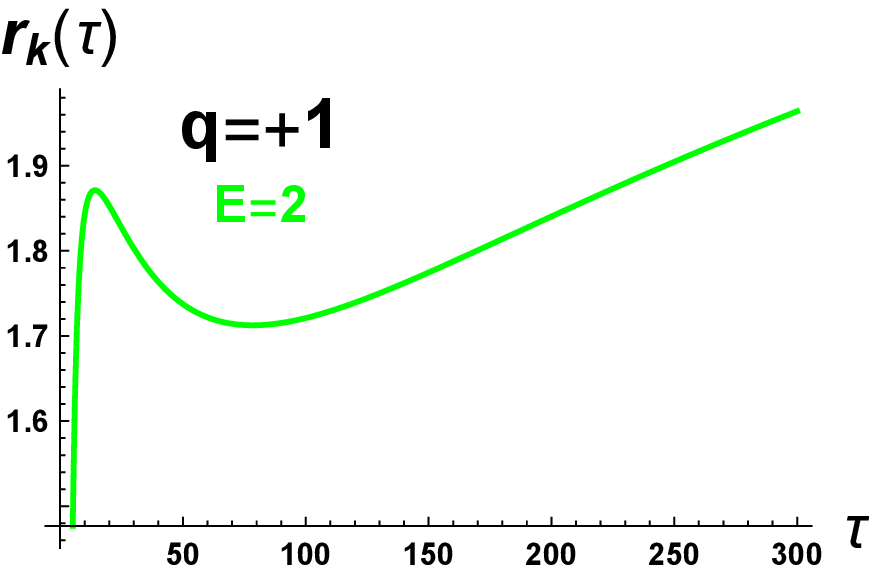}
		\includegraphics[scale=0.45]{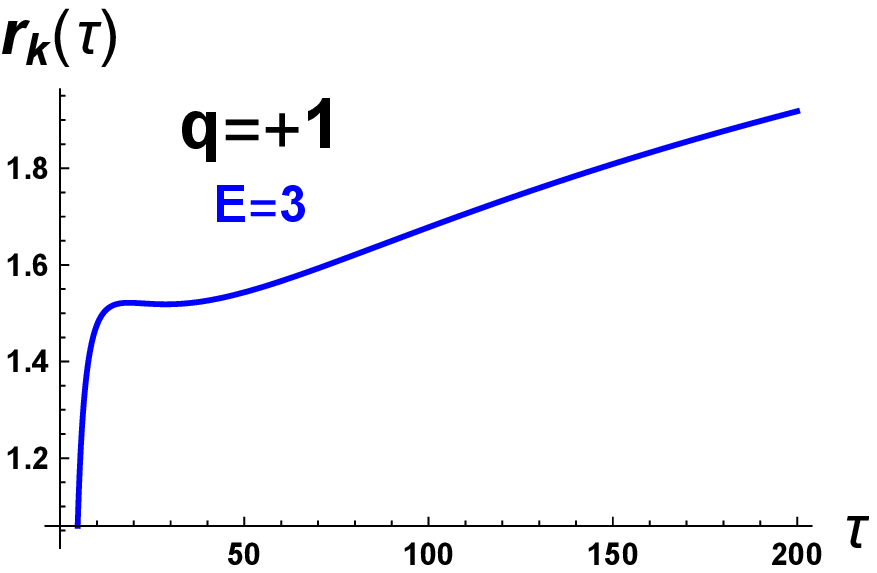}
		\caption{(color online). Variation of squeezing parameter $r_k$ with the brane proper time $\tau$ in BPS state ($q=1$) with different values of energy $E$.}
		\label{rkforBPstate}
	\end{center}
\end{figure}

\begin{figure}[ht]
	\begin{center}
		\includegraphics[scale=0.45]{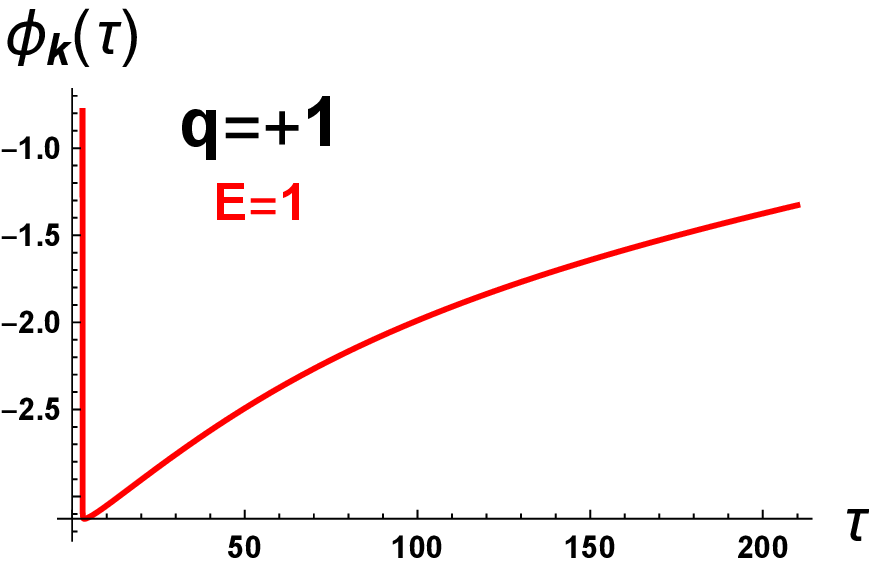}
		\includegraphics[scale=0.45]{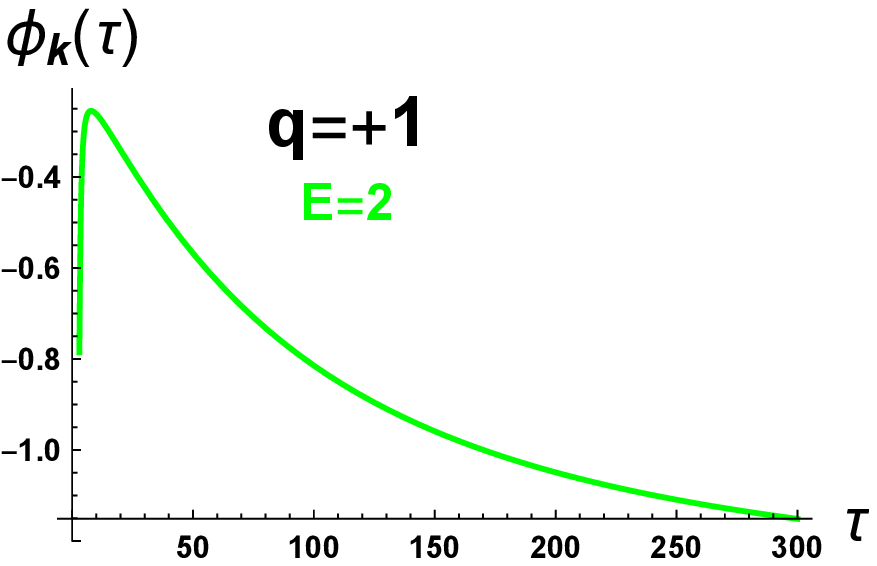}
		\includegraphics[scale=0.45]{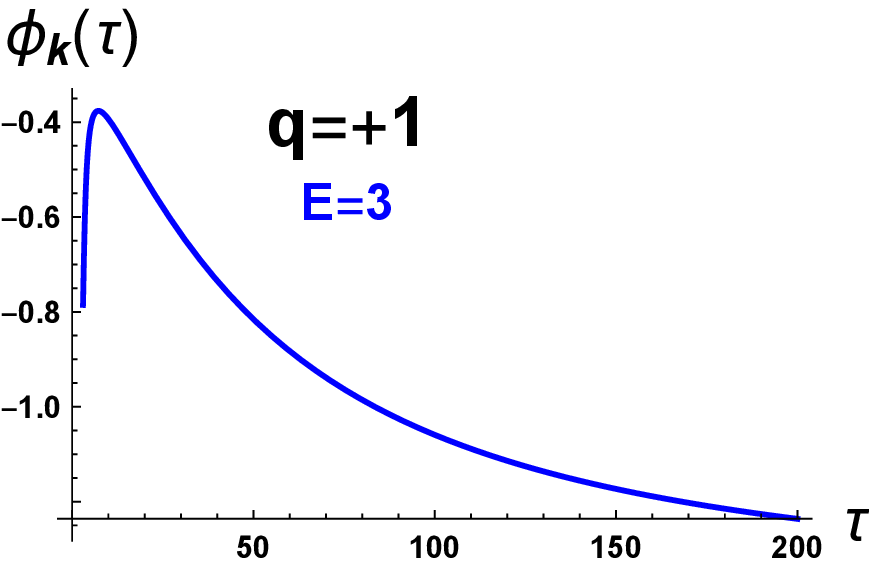}
		\caption{(color online). Variation of squeezing angle $phi_k$ with the brane proper time $\tau$ in BPS state ($q=1$) with different values of energy $E$.}
		\label{phikforBPstate}
	\end{center}
\end{figure}

\begin{figure}[ht]
	\begin{center}
		\includegraphics[scale=0.45]{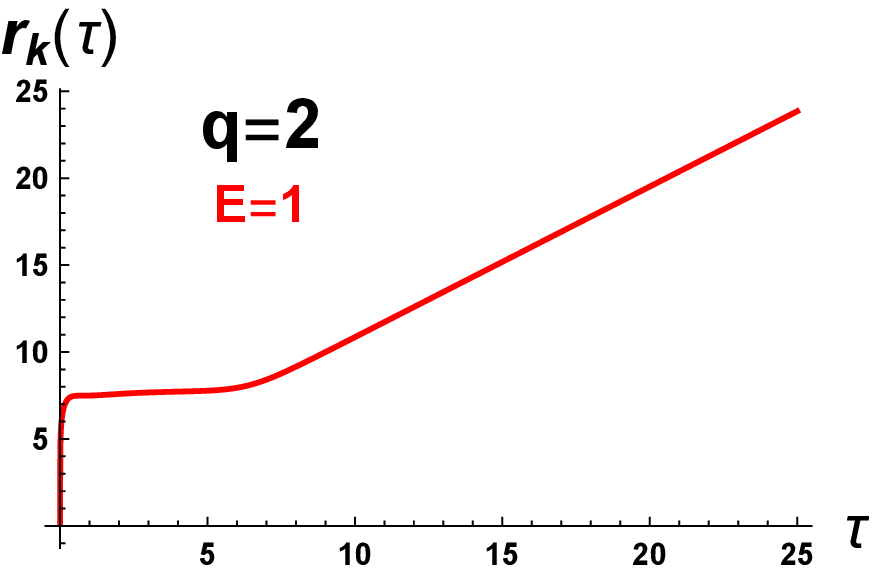}
		\includegraphics[scale=0.45]{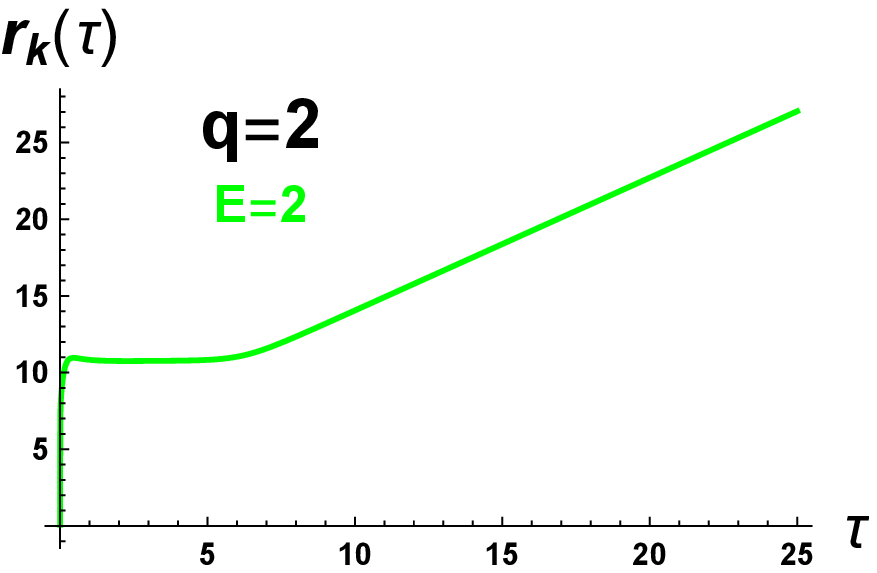}
		\includegraphics[scale=0.45]{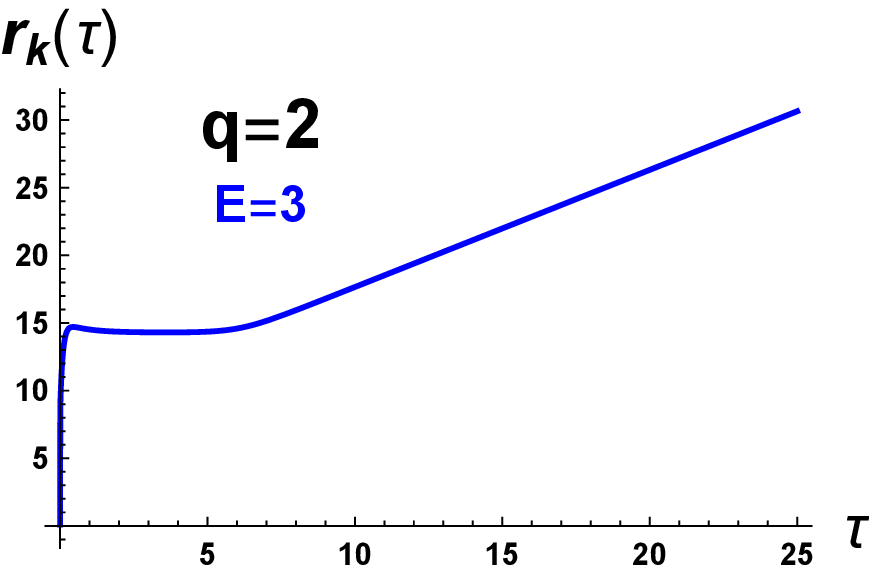}
		\caption{(color online). Variation of squeezing parameter $r_k$ with the brane proper time $\tau$ in non-BPS state ($q=2$) with different values of energy $E$.}
		\label{rkforABPstate}
	\end{center}
\end{figure}

\begin{figure}[ht]
	\begin{center}
		\includegraphics[scale=0.45]{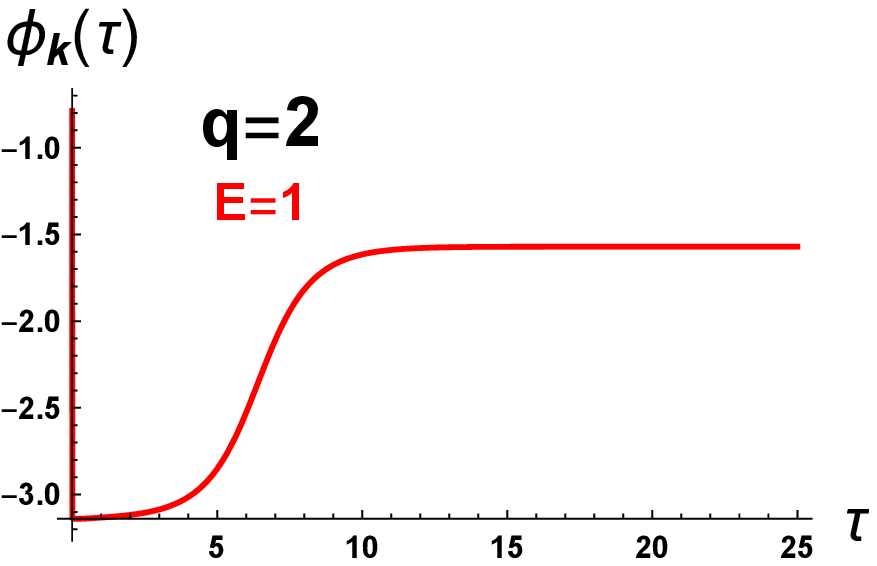}
		\includegraphics[scale=0.45]{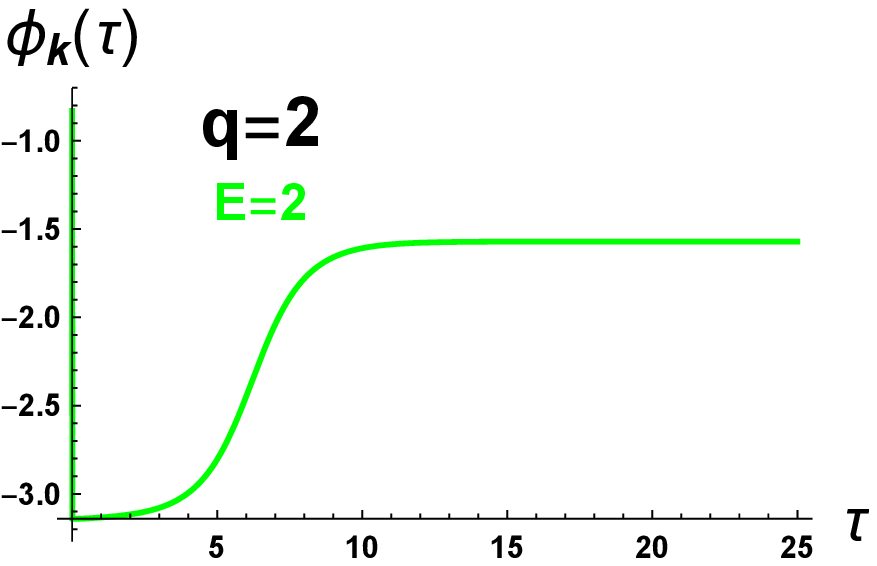}
		\includegraphics[scale=0.45]{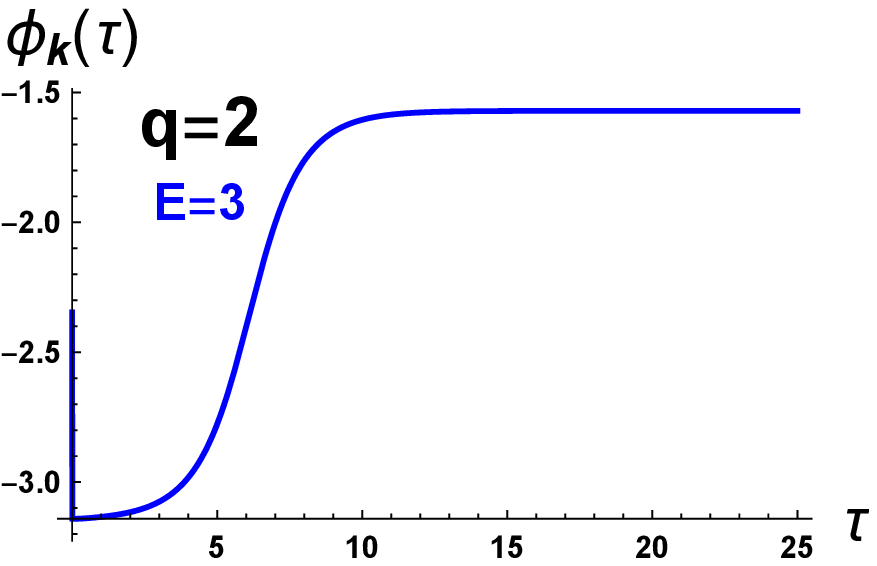}
		\caption{(color online). Variation of squeezing angle $\phi_k$ with the brane proper time $\tau$ in non-BPS state ($q=2$) with different values of energy $E$.}
		\label{phikforABPstate}
	\end{center}
\end{figure}

Analogous to the inverted harmonic oscillator case, the unitary evolution operator of a state could be factorized as
\begin{align}
&\hat{\mathcal{U}}_{\vec{k}}(\eta,\eta_{0})=\hat{\mathcal{S}}_{\vec{k}}(r_{k},\phi_{k})\hat{\mathcal{R}}_{\vec{k}}(\theta_{k})
\end{align}
in which the $\hat{\mathcal{R}}_{\vec{k}}(\theta_{k})$ represents the two-mode rotation operator characterized by rotation angle $\theta(\eta)$ 
\begin{align}
&\hat{\mathcal{R}}_{\vec{k}}(\theta_{k})=\exp\big[-i\theta_{k}(\eta)\big(\hat{c}_{\vec{k}}\hat{c}_{\vec{k}}^{\dagger}+\hat{c}_{-\vec{k}}^{\dagger}\hat{c}_{-\vec{k}}\big)\big]
\end{align}
And $\hat{\mathcal{S}}_{\vec{k}}$ is the two-mode squeeze operator defined by
\begin{align} 
\label{squOpe}
&\hat{\mathcal{S}}_{\vec{k}}(r_{k},\phi_{k})=\exp\big[r_{k}(\eta)\big(e^{-2i\phi_{k}(\eta)}\hat{c}_{\vec{k}}\hat{c}_{-\vec{k}}-e^{2i\phi_{k}(\eta)}\hat{c}_{-\vec{k}}^{\dagger}\hat{c}_{\vec{k}}^{\dagger}\big)\big]
\end{align}
where $r_k(\eta)$ and $\phi_k(\eta)$ are the squeezing parameter and squeezing angle respectively. The effects of rotation operator will be ignored since it only produces an irrelevant phase when acting on the initial vacuum state. By means of the ordering theorem of operator expansion \cite{TQO}, the $\eqref{squOpe}$ is rewritten as
\begin{align}
\nonumber
&\hat{\mathcal{S}}_{\vec{k}}(r_{k},\phi_{k})=\exp\big[-e^{2i\phi_{k}}\tanh r_{k}~\hat{c}_{-\vec{k}}^{\dagger}\hat{c}_{\vec{k}}^{\dagger}\big]\cdot\exp\big[-\ln(\cosh r_{k})\big(\hat{c}_{-\vec{k}}^{\dagger}\hat{c}_{-\vec{k}}+\hat{c}_{\vec{k}}\hat{c}_{\vec{k}}^{\dagger}\big)\big]\\
\label{squOpeV1}
&\quad\quad\quad\quad\quad\cdot\exp\big[e^{-2i\phi_{k}}\tanh r_{k}~\hat{c}_{\vec{k}}\hat{c}_{-\vec{k}}\big]
\end{align}
After acting operator $\eqref{squOpeV1}$ on the two-mode vacuum state $\vert 0; 0\rangle_{\vec{k},-\vec{k}}$, a two-mode squeezed state is constructed as
\begin{align}
\label{SquState}
&\vert\Psi\rangle_{sq}=\frac{1}{\cosh r_{k}}\big(\sum_{n=0}^{\infty}(-1)^{n}e^{-2in\phi_{k}}\tanh^{n}r_{k}\vert n;n\rangle_{\vec{k},-\vec{k}}\big)
\end{align}
where the two-mode excited state $\vert n;n\rangle_{\vec{k},-\vec{k}}$ has the expression
\begin{align}
&\vert n;n\rangle_{\vec{k},-\vec{k}}=\frac{1}{n!}\big(\hat{c}_{\vec{k}}^{\dagger}\big)^{n}\big(\hat{c}_{-\vec{k}}^{\dagger}\big)^{n}\vert0;0\rangle_{\vec{k},-\vec{k}}
\end{align}
The time evolution of $r_k(\eta)$ and $\phi_k(\eta)$ are controlled by the $Schr\ddot{o}dinger$ equation
\begin{align}
\label{EffSchroder}
&i\frac{d}{d\eta}\vert\Psi\rangle_{sq}=\hat{\mathcal{H}}_k \vert\Psi\rangle_{sq}
\end{align}
After substituting $\eqref{SqueeHamiltonian}$ and $\eqref{SquState}$ into $\eqref{EffSchroder}$, it yields
\begin{align}
\label{Solrketa}
&r_{k}^{\prime}=-\frac{a^{\prime}}{a}\cos2\phi_{k}-\frac{m^{2}(\eta)}{2k}\sin2\phi_{k}\\
\label{Solphiketa}
&\phi_{k}^{\prime}=(k+\frac{m^{2}(\eta)}{2k})+\frac{a^{\prime}}{a}\sin2\phi_{k}\coth2r_{k}-\frac{m^{2}(\eta)}{2k}\cos2\phi_{k}\coth2r_{k}
\end{align}
in which the prime denotes the derivative with respect to the conformal time $\eta$. It is convenient to work in terms of coordinate $\tau$, and hence the $\eqref{Solrketa}$-$\eqref{Solphiketa}$ are transformed to
\begin{align}
&\frac{dr}{d\tau}=-\frac{\dot{a}}{a}\cos2\phi_{k}-\frac{m^{2}(\tau)}{2k}\frac{\sin2\phi_{k}}{a}\\
&\frac{d\phi_{k}}{d\tau}=\frac{1}{a}(k+\frac{m^{2}(\tau)}{2k})+\frac{\dot{a}}{a}\sin2\phi_{k}\coth2r_{k}-\frac{m^{2}(\tau)}{2k}\frac{\cos2\phi_{k}\coth2r_{k}}{a}
\end{align}
After plugging the numerical solutions given by Fig.\ref{SolFriedInBPS} into the above differential equations, the evolution of $r_k$ and $\phi_k$ in different $E$ with $q=1,2$ respectively are shown in Fig.\ref{rkforBPstate}-Fig.\ref{phikforABPstate}. For BPS state depicted in Fig.\ref{rkforBPstate}-Fig.\ref{phikforBPstate}, as the value of $E$ is increased, the squeezing parameter $r_k(\tau)$ will be monotonically increasing and the inflection point will disappear. Meanwhile, the trend of the change of the squeezing angle $\phi_k$ with higher $E$ is opposite to the one with lower $E$. Note that significant difference between the case of $E=1$ and the case of $E=2,3$ are originated from the $m^2(\tau)$, as shown by the left part in Fig.\ref{EffecSquMass}. In regard to non-BPS state, as shown in Fig.\ref{rkforABPstate}-Fig.\ref{phikforABPstate}, the variation tendency of $r_k$ and $\phi_k$ will not change if we increase the values of $E$.
In Fig.\ref{rkforABPstate}, we notice that the $r_k$ with higher $E$ will enter into the linear growth regime in earlier time. From the Fig.\ref{phikforABPstate}, we find that the $\phi_k$ with higher $E$ will approach a constant more quickly.

\subsection{The quantum circuit complexity}

\begin{figure}[ht]
	\begin{center}
		\includegraphics[scale=0.45]{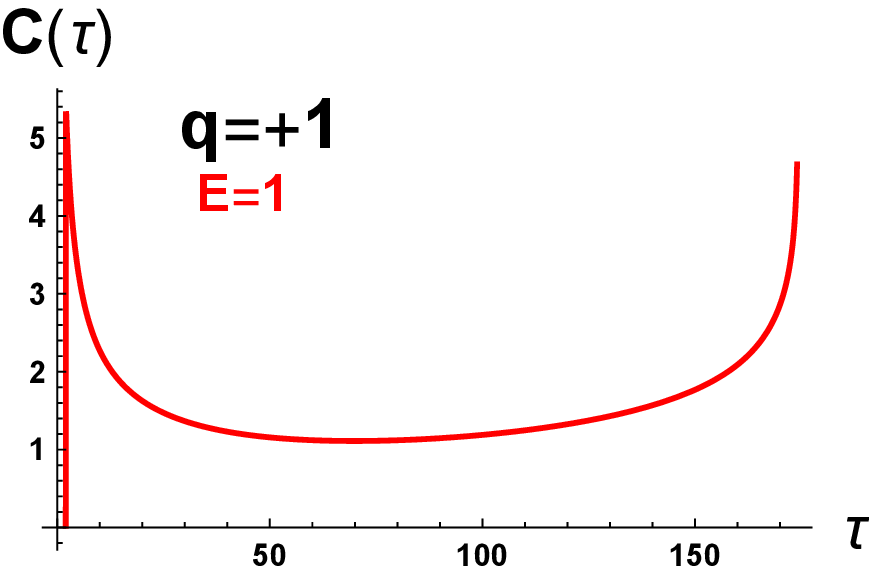}
		\includegraphics[scale=0.45]{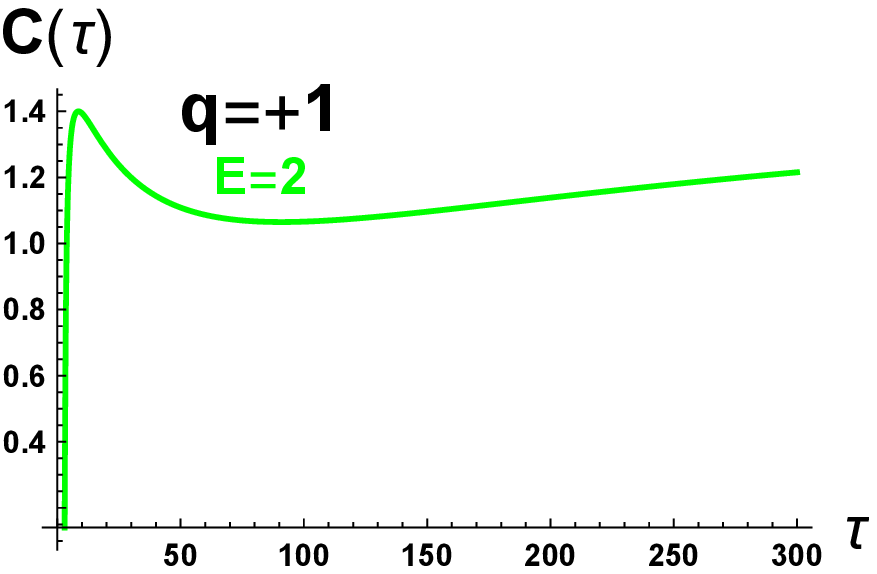}
		\includegraphics[scale=0.45]{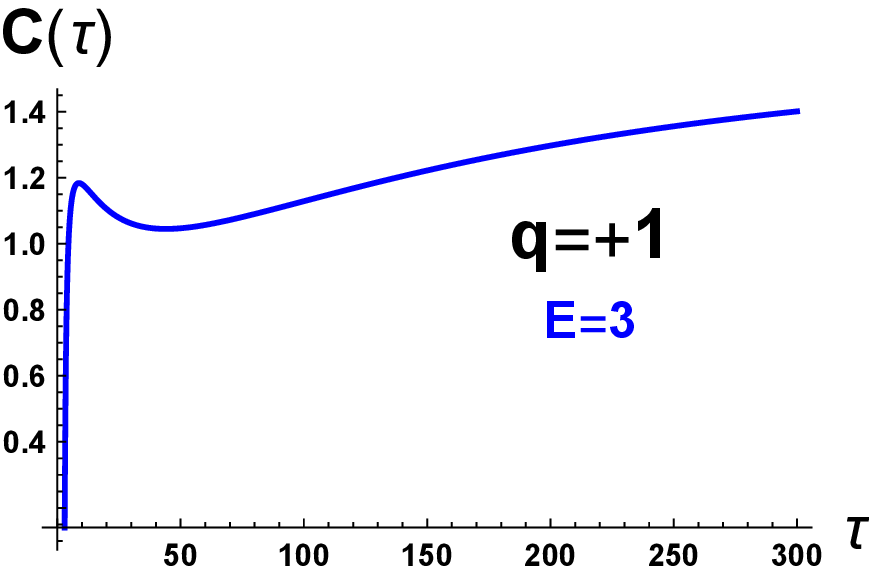}
		\caption{(color online). The evolution of complexity as the brane proper time $\tau$ increases in BPS state (q=1) with different values of energy $E$.}
		\label{ComBPSstate}
	\end{center}
\end{figure}

\begin{figure}[ht]
	\begin{center}
		\includegraphics[scale=0.45]{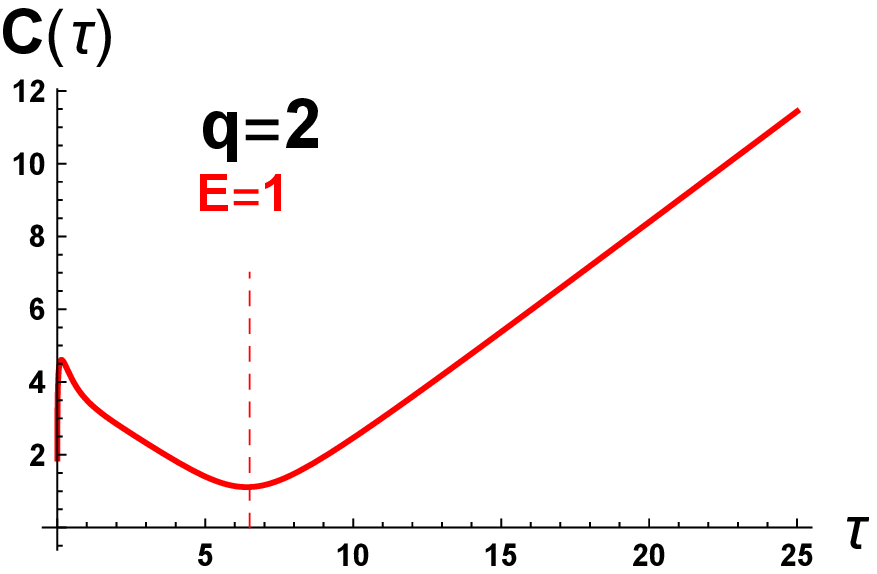}
		\includegraphics[scale=0.45]{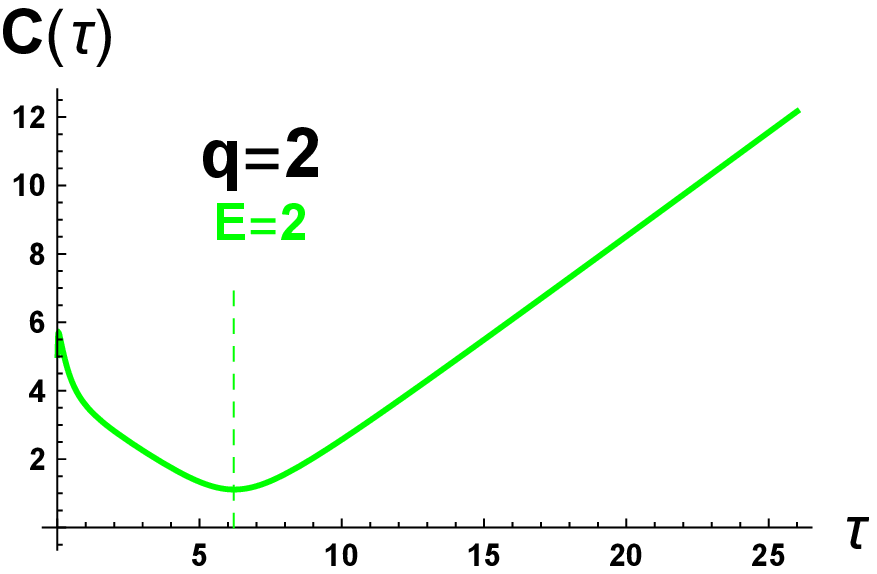}
		\includegraphics[scale=0.45]{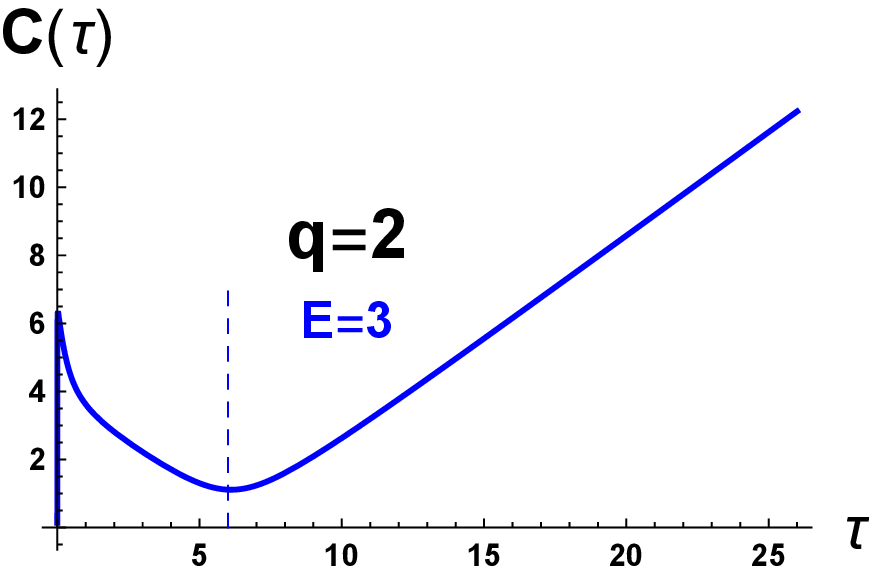}
		\caption{(color online). The evolution of complexity as the brane proper time $\tau$ increases in anti-BPS state (q=2) with different values of energy $E$.}
		\label{ComABPSstate}
	\end{center}
\end{figure}

In this paper, Nielsen's method \cite{NielsenComplexity1, NielsenComplexity2, NielsenComplexity3} are used to evaluate the quantum circuit complexity. Given a reference state $\vert \psi^R\rangle$ at $\tau=0$, it will evolve into a target state $\vert \psi^T \rangle$ through an unitary operator, namely
\begin{align}
\label{ComWaveFunc}
&\vert \psi ^T \rangle_{\tau=1}=U(\tau=1) \vert \psi ^R \rangle_{\tau=0}
\end{align}
Notice that $\tau$ parametrizes a path in the Hilbert space. In particular, one could construct the unitary operator from a path-ordered exponential of a Hamiltonian operator
\begin{align}
&U(\tau)=\overleftarrow{\mathcal{P}} \exp \bigg( -i\int ^\tau _0 ds H(s) \bigg)
\end{align}
where the $\overleftarrow{\mathcal{P}}$ represents right-to-left path ordering. In general, the Hamiltonian operator $H(\tau)$ can be decomposed into a basis of Hermitian operators $M_I$, which are the generators for elementary quantum gates
\begin{align}
&H(\tau)= Y(\tau)^I M_I
\end{align}
As control function, $Y(\tau)^I$ determine which quantum gate should be switched on or off at a fixed parameter $\tau$. The evolution of the unitary operator obeys the $Schr\ddot{o}dinger$ equation
\begin{align}
&\frac{dU(\tau)}{d\tau}=-iY(\tau)^I M_I U(\tau)
\end{align}
And then one could define a $cost~functional$ 
\begin{align}
\label{costfunctional}
&\mathcal{C}(U)=\int_0 ^1 \mathcal{F(U,\dot{U})}d\tau
\end{align}
in which the $\mathcal{F}$ is generally based upon the control functions $Y(\tau)^I$. The complexity is obtained by minimizing the functional $\eqref{costfunctional}$, as shown by \cite{Chapman:2017rqy,Jefferson:2017sdb}, which is equivalent to the operation that find the shortest geodesic distance between the reference and target states in operator space. Specifically, we restrict our attentions on the $quadratic$ cost functional \cite{Bhattacharyya:2020rpy}
\begin{align}
\label{quadfunctional}
&\mathcal{F}(U,Y)=\sqrt{\sum_I (Y^I)^2}
\end{align}
In our consideration, the target state is the two-mode squeezed vacuum state $\eqref{SquState}$. After projecting $\vert\Psi\rangle$ into the position space, the following wavefunction is constructed \cite{Martin:2019wta, Lvovsky:2014sxa}
\begin{align}
\nonumber
&\Psi_{sq}(q_{\vec{k}},q_{-\vec{k}})=\sum_{n=0}^{\infty}(-1)^{n}e^{2in\phi_{k}}\frac{\tanh^{n}r_{k}}{\cosh r_{k}}\langle q_{\vec{k}};q_{-\vec{k}}\vert n;n\rangle_{\vec{k},-\vec{k}}\\
\label{squeeposition}
&\quad\quad\quad\quad\quad~=\frac{\exp[A(r_{k},\phi_{k})\cdot(q_{\vec{k}}^{2}+q_{-\vec{k}}^{2})-B(r_{k},\phi_{k})\cdot q_{\vec{k}}q_{-\vec{k}}]}{\cosh r_{k}\sqrt{\pi}\sqrt{1-e^{-4i\phi_{k}}\tanh^{2}r_{k}}}
\end{align}
where the coefficients $A(r_k,\phi_k)$ and $B(r_k,\phi_k)$ have the expressions
\begin{align}
&A(r_{k},\phi_{k})=\frac{k}{2}\bigg(\frac{e^{-4i\phi_{k}}\tanh^{2}r_{k}+1}{e^{-4i\phi_{k}}\tanh^{2}r_{k}-1}\bigg)\\
&B(r_{k},\phi_{k})=2k\bigg(\frac{e^{-2i\phi_{k}}\tanh r_{k}}{e^{-4i\phi_{k}}\tanh^{2}r_{k}-1}\bigg)
\end{align}
In term of the matrix form, the exponent in $\eqref{squeeposition}$ can be transformed to a diagonal one by acting an appropriate rotation operation in vector space $(q_{\vec{k}},q_{-\vec{k}})$
\begin{align}
\label{targerstate}
&\Psi_{sq}(q_{\vec{k}},q_{-\vec{k}})=\frac{\exp[-\frac{1}{2}\tilde{M}^{ab}q_{a}q_{b}]}{\cosh r_{k}\sqrt{\pi}\sqrt{1-e^{-4i\phi_{k}}\tanh^{2}r_{k}}}
\end{align}
where
\begin{align}
\nonumber
&\tilde{M}=\left(\begin{array}{cc}
\Omega_{\vec{k}^{\prime}} & 0\\
0 & \Omega_{-\vec{k}^{\prime}}
\end{array}\right)=\left(\begin{array}{cc}
-2A+B & 0\\
0 & -2A-B
\end{array}\right)
\end{align}
At the same time, the reference state is the unsqueezed vacuum state
\begin{align}
\nonumber
&\Psi_{00}(q_{\vec{k}},q_{-\vec{k}})=\langle q_{\vec{k}};q_{-\vec{k}}\vert0;0\rangle_{\vec{k},-\vec{k}}\\
\nonumber
&\quad\quad\quad\quad\quad~=\frac{\exp[-\frac{1}{2}(\omega_{\vec{k}}q_{\vec{k}}^{2}+\omega_{-\vec{k}}q_{-\vec{k}}^{2})]}{\pi^{1/2}}\\	
\label{referstate}	
&\quad\quad\quad\quad\quad~=\frac{\exp[-\frac{1}{2}\tilde{m}^{ab}q_aq_b]}{\pi^{1/2}}
\end{align}
in which
\begin{align}
&\tilde{m}=\left(\begin{array}{cc}
\omega_{\vec{k}} & 0\\
0 & \omega_{-\vec{k}}
\end{array}\right)=\left(\begin{array}{cc}
\vert\vec{k}\vert & 0\\
0 & \vert-\vec{k}\vert
\end{array}\right)=\left(\begin{array}{cc}
k & 0\\
0 & k
\end{array}\right)
\end{align}
For a target and a reference state wavefunction $\eqref{targerstate}$-$\eqref{referstate}$ in matrix form, the relation $\eqref{ComWaveFunc}$ is rewritten as 
\begin{align}
\label{Utrans1}
&\Psi_\tau(q_{\vec{k}},q_{-\vec{k}})=\tilde{U}(\tau)\Psi_{00}(q_{\vec{k}},q_{-\vec{k}}) \tilde{U}^\dagger (\tau)\\
\label{boun1}
&\Psi_{\tau=0}(q_{\vec{k}},q_{-\vec{k}})=\Psi_{00}(q_{\vec{k}},q_{-\vec{k}})\\
\label{boun2}
&\Psi_{\tau=1}(q_{\vec{k}},q_{-\vec{k}})=\Psi_{sq}(q_{\vec{k}},q_{-\vec{k}})
\end{align}
where $\tilde{U}^\dagger (\tau)$ represents the general linear group in two dimension complex space, i.e.$GL(2,C)$. With the aid of Lie algebra, $\tilde{U} (\tau)$ has the expression
\begin{align}
&\tilde{U}(\tau)=\exp[\sum_{I=1}^{4}Y^I (\tau) M_I]
\end{align}
where the $\{M_I\}$ are the 4 generators of $GL(2,C)$, namely
\begin{align}
&M_1=\left(\begin{array}{cc}
1 & 0\\
0 & 0
\end{array}\right)~,~M_2=\left(\begin{array}{cc}
0 & 0\\
0 & 1
\end{array}\right)~,~
M_3=\left(\begin{array}{cc}
0 & 1\\
0 & 0
\end{array}\right)~,~M_4=\left(\begin{array}{cc}
0 & 0\\
1 & 0
\end{array}\right)
\end{align}
Since the manifold of general linear group corresponds to an Euclidean geometry in complex space, the geodesic distance in operator space of $GL(N,C)$ is characterized by the line element
\begin{align}
&ds^2=\delta_{IJ}dY^I d{(Y^J)^{\star}}
\end{align} 
In language of geometry, the complexity $\eqref{costfunctional}$ together with the quadratic cost functional is rewritten by the following line length
\begin{align}
\label{CompleLineLeng}
&C(\tilde{U})=\int^1_0 d\tau \sqrt{G_{IJ}\dot{Y}^I(\tau) ({\dot{Y}}^J(\tau))^\star}
\end{align}
where the dot denotes the derivative with respect to the $\tau$. It is easy to see that the shortest geodesic between target state and reference in geometry space of $GL(2,C)$ is just a straight line
\begin{align}
\label{strailineYtau}
&Y^I (\tau)=Y^I (\tau=1)\cdot \tau+ Y^I (\tau=0)
\end{align}
In $\eqref{Utrans1}$, since both the $\Psi_{00}(q_{\vec{k}},q_{-\vec{k}})$ and $\Psi_{sq}(q_{\vec{k}},q_{-\vec{k}})$ are in diagonal matrix form, the off-diagonal generators will increase the distance between two states in operator. And hence components $Y^3$ and $Y^4$ are set to be zero \cite{Jefferson:2017sdb}. The equation $\eqref{strailineYtau}$ and boundary conditions $\eqref{boun1}$-$\eqref{boun2}$ mean that the $Y^{1,2}$ must satisfy
\begin{align}
\label{Ytau0}
&\quad~\text{Im}(Y^{1,2})\big\vert_{\tau=0}=\text{Re}(Y^I)\big\vert_{\tau=0}=0\\
\label{Ytau1}
&\begin{cases}
\begin{array}{c}
\hspace{-10mm}\text{Im}(Y^{1,2})\big\vert_{\tau=1}=\frac{1}{2}\ln\frac{\vert\Omega_{\vec{k},\vec{-k}}\vert}{\omega_{\vec{k},\vec{-k}}}\\
\text{Re}(Y^{1,2})\big\vert_{\tau=1}=\frac{1}{2}\arctan\frac{\text{Im}(\Omega_{\vec{k},\vec{-k}})}{\text{Re}(\Omega_{\vec{k},\vec{-k}})}
\end{array}\end{cases}
\end{align}
After plugging $\eqref{strailineYtau}$-$\eqref{Ytau1}$ into $\eqref{CompleLineLeng}$, the complexity has the expression
\begin{align}
\label{CacComplexity}
&\mathcal{C}(k)=\frac{1}{2}\sqrt{\big(\ln\frac{\vert\Omega_{\vec{k}}\vert}{\omega_{\vec{k}}}\big)^{2}+\big(\ln\frac{\vert\Omega_{-\vec{k}}\vert}{\omega_{-\vec{k}}}\big)^{2}+\big(\arctan\frac{\text{Im}(\Omega_{\vec{k}})}{\text{Re}(\Omega_{\vec{k}})}\big)^{2}+\big(\arctan\frac{\text{Im}(\Omega_{\vec{-k}})}{\text{Re}(\Omega_{\vec{-k}})}\big)^{2}}
\end{align}
The evolution of complexity is obtained in Fig.\ref{ComBPSstate}-Fig.\ref{ComABPSstate} by substituting the numerical solutions of $r_k$ and $\phi_k$ in expression $\eqref{CacComplexity}$. In case of non-BPS brane, as shown by Fig.\ref{ComABPSstate}, it is easy to observe the linear growth behavior of the complexity, which means the appearance of the chaos \cite{Ali:2019zcj}. In particular, the time scale when the complexity starts to grow (as denoted by the dotted line in picture) could be identified as the scrambling time, while the Lyapunov exponent approximately equals to the slope of the linear growth part. And the system will evolve into the chaotic regime (linear growth regime) more earlier as we improved the values of energy $E$. However, in case of BPS brane, there does not appear the chaotic behavior distinctly. To be more exact, it is necessary to consider the corresponding OTOCs for the Fig.\ref{ComBPSstate}. And we will leave it as the future works.

\section{Conclusions and Discussion}

In this paper, we use geometric method developed in areas of quantum information to consider the quantum circuit complexity between initial vacuum state and the squeezed quantum state evolved by the quantum fluctuations around the probe $D_3$ brane moving in black 3-brane background. For simplification, we mainly focus on a $AdS_5$-Schwarzchild bulk which corresponds to the near horizon limit of the black 3-brane geometry. We consider the embedding of $D_3$ brane worldvolume in bulk geometry and obtain a Friedmann-like equation from the conservation law of energy. And then, the quadratic perturbative action describing the quantum fluctuations in scalar mode around the probe brane is constructed by expanding the DBI action in term of diffeomorphism invariant measure of infinisimal spacetime displacement. The quadratic perturbative action could be simplified further by utilizing the $Gauss$-$Codacci$ equations. The squeezed state is obtained by acting the two-mode squeeze operator on vacuum state, which is characterized by squeezing parameters $r_k$ and $\phi_k$. The differential equations governing the evolution of squeezing parameters are acquired by acting the Hamiltonian operator derived from the perturbative action on squeezed  state. The numerical solutions of $r_k(\tau)$ and $\phi_k(\tau)$ for the BPS brane and non-BPS brane are displayed in Fig.\ref{rkforBPstate}-Fig.\ref{phikforABPstate}. 

Finally, the wave-function approach is used to evaluate the quantum circuit complexity of the quantum fluctuations around the moving probe $D_3$ brane, in which the reference state and the target state are chosen as vacuum state and squeezed state respectively. As a diagnostic for quantum chaos, the evolution of quantum circuit complexity given in Fig.\ref{ComBPSstate}-Fig,\ref{ComABPSstate} reveal that the quantum fluctuations around the non-BPS brane manifestly evolve into the chaotic regime at the late time, while the chaotic behavior is not easy to observe in case of BPS brane. From the viewpoints of holography, it implies that the thermodynamic system consist of the $\mathcal{N}=4$ supersymmetric particles in non-BPS states evolve into a chaotic system more easily than the one in BPS states. This implication is accepted in intuition since the BPS states are the lightest charged particles and are more stable than non-BPS states. 

However, an limitation of this work is that the perturbative action of probe $D_3$ brane is considered only up to quadratic orders. Thus, an interesting work is to explore the effects of higher-order perturbations of DBI action on the evolution of quantum circuit complexity. On the other hand, although the quantum circuit complexity could be viewed as a diagnostic tool to inspect the quantum chaos, on could also reexamine the results in this work by using other methods such as the out-of-time-ordered correlator. Besides, if we consider the motion of probe brane with negative brane charges, it will move inward to black hole and fall into the horizon at the end. It is an interesting topic to consider the evolution of squeezed quantum states after the brane entering into the interior of black hole, which could be associated to the black hole information paradox.

\end{document}